\documentclass[journal]{IEEEtran}
\usepackage{cite}
\usepackage{graphicx}
\usepackage{subfigure}
\usepackage{algorithm} 
\usepackage{algorithmic}

\begin{document}
%
% paper title
% Titles are generally capitalized except for words such as a, an, and, as,
% at, but, by, for, in, nor, of, on, or, the, to and up, which are usually
% not capitalized unless they are the first or last word of the title.
% Linebreaks \\ can be used within to get better formatting as desired.
% Do not put math or special symbols in the title.
\title{Deployment Strategies of Multiple Aerial BSs for User Coverage and Power Efficiency Maximization}
%
%
% author names and IEEE memberships
% note positions of commas and nonbreaking spaces ( ~ ) LaTeX will not break
% a structure at a ~ so this keeps an author's name from being broken across
% two lines.
% use \thanks{} to gain access to the first footnote area
% a separate \thanks must be used for each paragraph as LaTeX2e's \thanks
% was not built to handle multiple paragraphs
%

\author{Jingcong~Sun,~\IEEEmembership{Student Member,~IEEE,}
        Christos~Masouros,~\IEEEmembership{Senior Member,~IEEE}\thanks{J.Sun and C.Masouros are with the Department of Electronic and Electrical
Engineering,  University  College  London,  London  WC1E  7JE,  U.K.  (e-mail: uceejsu@ucl.ac.uk; chris.masouros@ieee.org)}}

\maketitle

% As a general rule, do not put math, special symbols or citations
% in the abstract or keywords.
\begin{abstract}
Unmanned aerial vehicle (UAV) based aerial base stations (BSs) can provide rapid communication services to ground users and are thus promising for future communication systems. In this paper, we consider a scenario where no functional terrestrial BSs are available and the aim is deploying multiple aerial BSs to cover a maximum number of users within a certain target area. To this end, we first propose a naive successive deployment method, which converts the non-convex constraints in the involved optimization into a combination of linear constraints through geometrical relaxation. Then we investigate a deployment method based on $K$-means clustering. The method divides the target area into $K$ convex subareas, where within each subarea, a mixed integer non-linear problem (MINLP) is solved. An iterative power efficient technique is further proposed to improve coverage probability with reduced power. Finally, we propose a robust technique for compensating the loss of coverage probability in the existence of inaccurate user location information (ULI). Our simulation results show that, the proposed techniques achieve an up to 30\% higher coverage probability when users are not distributed uniformly. In addition, the proposed simultaneous deployment techniques, especially the one using iterative algorithm improve power-efficiency by up to 15\% compared to the benchmark circle packing theory. 
\end{abstract}

% Note that keywords are not normally used for peerreview papers.
\begin{IEEEkeywords}
Unmanned aerial vehicles, user coverage, air-to-ground communication, clustering algorithm
\end{IEEEkeywords}

% For peer review papers, you can put extra information on the cover
% page as needed:
% \ifCLASSOPTIONpeerreview
% \begin{center} \bfseries EDICS Category: 3-BBND \end{center}
% \fi
%
% For peerreview papers, this IEEEtran command inserts a page break and
% creates the second title. It will be ignored for other modes.
\IEEEpeerreviewmaketitle

\section{Introduction}
% The very first letter is a 2 line initial drop letter followed
% by the rest of the first word in caps.
% 
% form to use if the first word consists of a single letter:
% \IEEEPARstart{A}{demo} file is ....
% 
% form to use if you need the single drop letter followed by
% normal text (unknown if ever used by the IEEE):
% \IEEEPARstart{A}{}demo file is ....
% 
% Some journals put the first two words in caps:
% \IEEEPARstart{T}{his demo} file is ....
% 
% Here we have the typical use of a "T" for an initial drop letter
% and "HIS" in caps to complete the first word.
\IEEEPARstart{L}{ow-altitude} unmanned aerial vehicles (UAVs) have been increasingly appealing to future wireless communication systems. UAVs which are cost-effective, interoperable, flexible and likely to have a higher probability of line-of-sight (LoS) channels are promising in a various scenarios for both civilian and military use \cite{Namuduri:2013:MAH:2491260.2491265,4745652,7122576}. UAV-based communication applications involve three main categories, namely relaying, information dissemination/data collection and ubiquitous coverage \cite{7470933}. UAVs serving as relaying nodes are studied to provide reliable wireless communication between distant users with blocked direct links while increasing the system throughput \cite{4225050,7572068,8068199,8170970}. Optimizing the flying trajectory is of particular interest when UAVs are dispatched to disseminate information \cite{5700250,7888557}. Last but not least, as UAVs can be quickly deployed, UAV-based aerial base stations (BSs) are attracting increasing interest as a means to provide fast wireless services to ground users. 
For instance, UAVs can be deployed to ease the burden of terrestrial base stations in extremely crowded areas by offloading users from ground cells when specific rate or distance requirement is satisfied \cite{8292783,7461487}. Moreover, fast service recovery or supply can be offered by such flying BSs when fixed communication infrastructures are damaged. 
\\\indent With the rising interest in the above areas, the challenges in the practical use of aerial BSs are becoming pertinent. When the aerial BSs are utilized for emergency communications such as search-and-rescue, the priority is finding the optimal locations of UAVs so that a maximum number of users can be covered. Meanwhile, since built-in batteries are used for supplying power in most cases, limited on-board energy is another factor that constrains the endurance of aerial BSs \cite{7470933,8304077,8316986,8269064}. It has been proven that prolonged operation time can be achieved by reducing the transmit power of aerial BSs when quality-of-service (QoS) requirements are met \cite{7317490,7918510}. 
\\\indent The aerial BS coverage problem is first studied in \cite{6863654}, which gave an air-to-ground (ATG) channel model used to find the optimal altitude of UAVs that can lead to maximum coverage area on the ground. Instead of just maximizing the coverage area, recent research has become increasingly focused on algorithms trying to cover the maximum number of users \cite{7918510,8038014,7510820,7962642}. Specifically, \cite{7918510} solved a 3-D circle placement problem by formulating it as a mixed integer non-linear problem (MINLP), while \cite{8038014} made a further step by considering explicit QoS constraints. In \cite{7962642}, optimal location of an aerial BS is obtained through an exhaustive search in predefined girds. However, all the works mentioned above considered only the case of a single flying BS which limits their use. In most real situations, it is necessary to deploy multiple UAVs at the same time to cover a majority of users in a specific target region. The work in \cite{7417609} extends the number of utilized UAVs to two with a careful consideration of inter-cell interference (ICI). Mozaffari \emph{et al.}\cite{7486987} proposed a circle packing technique so that the total area covered by multiple aerial BSs is maximized, however the method did not consider user distributions. A 100\% user coverage probability is shown through a spiral algorithm in \cite{7762053}. However, the study ignores the effect of ICI, and the interference issue needs to be tackled with overlaid techniques.\\
\indent  In this paper, we study the efficient deployment of multiple UAVs so the maximum user coverage probability is achieved, where we define the coverage probability as the ratio of number of covered users to the total number of users within a specific target area. Following \cite{7486987,7762053}, we assume that the locations of users are known with the help of high-accuracy GPS systems and each aerial BS has enough capacity to supply all the users it covers. We consider a scenario where multiple UAVs are deployed in a target area without ground BSs' coverage. This is relevant in rural area coverage in cases where terrestrial BSs are absent, and in natural disaster scenarios where terrestrial coverage is disabled. Rotary-wing UAVs which have the ability to move in arbitrary direction as well as hold still in the air are assumed as the carrier for aerial BSs \cite{7470933}.
Our aim is to study the efficient placement of multiple aerial BSs in order to obtain a maximum user coverage probability while completely avoiding the influence of ICI. The UAV placement problem is modelled as a circle placement problem and no coverage overlap is allowed so that ICI between aerial BSs is intrinsically avoided. Our simulation results demonstrate that the proposed circle placement methods achieve higher user coverage probability than the benchmark circle packing theory (CPT) \cite{7486987}. Moreover, the increased coverage probability is achieved with significantly reduced transmit power in certain scenarios. The existence of inaccurate user location information (ULI) is also considered, and clearly increased robustness against inaccurate ULI is obtained when proposed robust technique is applied.\\
\indent For clarity, We summarize the main contributions of this paper as follows\\
\begin{itemize}
    \item Geometrical Relaxation: we propose a geometrical relaxation scheme where the optimal locations of UAVs are obtained in a 'step-by-step' fashion, in which the next UAV is always deployed in a position such that it covers the most number of remaining users in the target area until there is no space for accommodating more UAVs. The formulation of such a problem includes a increased number of non-convex constraints to avoid interference between any two aerial BSs. The non-convex constraints are addressed with a simple geometrical relaxation which converts each non-convex constraint into four linear constraints that can be easily solved.
    \item $K$-means Deployment: a more efficient technique which can be easily applied to any size of target area is proposed with the help of $K$-means clustering algorithm \cite{Xu2015}, where the best locations of aerial BSs are found within several subareas which are convex regions.
    \item Power Efficient $K$-means Deployment: we then propose an iterative algorithm to further improve the user coverage probability while drastically reducing the required transmit power of aerial BSs.
    \item Robust Deployment with imperfect ULI: the coverage performance in the existence of imperfect ULI is also considered, and a robust technique to compensate the performance loss with minimum transmit power is proposed accordingly by shifting the location of aerial BSs within a bounded circular region before increasing the radii of corresponding coverage areas.
    \item Complexity Analysis: we derive the computational complexity of the proposed techniques analytically in terms of the floating-point operations required. It is shown mathematically that the complexity of the $K$-means deployment scales linearly with the number of UAVs and the number of users, while an additional quadratic scaling with the number of UAVs is observed for the robust techniques.
\end{itemize}
\vspace*{12 pt}

\indent The remainder of this paper is organized as follows. Section \uppercase\expandafter{\romannumeral2} introduces system model. The successive deployment method based on geometrical relaxation is proposed in Section \uppercase\expandafter{\romannumeral3}. Section \uppercase\expandafter{\romannumeral4} and Section \uppercase\expandafter{\romannumeral5} propose two simultaneous deployment methods based on $K$-means clustering, and Section \uppercase\expandafter{\romannumeral6} considers the situation of inaccurate ULI. Computational complexity of the proposed techniques are analyzed in Section \uppercase\expandafter{\romannumeral7} and numerical results are shown in Section \uppercase\expandafter{\romannumeral8}. Finally, the paper is concluded in Section \uppercase\expandafter{\romannumeral9}.

%\subsection{Subsection Heading Here}
%Subsection text here.

% needed in second column of first page if using \IEEEpubid
%\IEEEpubidadjcol

%\subsubsection{Subsubsection Heading Here}
%Subsubsection text here.

\section{System model}

We consider a square geographical target area with side length ${L_s}$ containing a set of low-mobility users denoted by $\cal{M}$ as shown in Fig. 1. We assume $K$ multiple aerial BSs are deployed within the region in order to provide wireless communication to as many ground users as possible. Note that, as UAVs can move freely, such deployment of aerial BSs can be done regularly in order to accommodate any changes in the user positions. We note that trajectory design is out of the scope of this paper and we will only focus on each snapshot of users within the area.
\subsection{Path loss model}
There are a few different models characterizing air-to-ground (AtG) links. Several papers simply use LoS channels as the AtG channel model due to the fact that the UAV-ground communication is more likely to be dominated by LoS links compared to terrestrial communication systems, especially in areas without a large number of high-risers such as suburban and rural areas \cite{7888557,7762053}.  Here, we utilize a more strict and general channel model as in \cite{6863654} which considers both the effect of LoS links and non-line-of-sight (NLoS) links. If we denote the location of user $i$ in the set $\cal{M}$ as $({x_i},{y_i})$, the horizontal location and the altitude of the $k$-th UAV as $({x_{ck}},{y_{ck}})$ and ${h_k}$, $k = 1,2,...K$ respectively, then the ground distance between user $i$ and UAV $k$ is ${r_{ik}} = \sqrt {{{({x_i} - {x_{ck}})}^2} + {{({y_i} - {y_{ck}})}^2}} $. Accordingly, the probability of having a LoS link is:
\begin{equation}
   {P_{{\rm{LoS}}}} = \frac{1}{{1 + a{\rm{ exp(}} - b{\rm{(}}\frac{{180}}{\pi }{\theta _{ik}} - a))}}
\end{equation}
where $a$ and $b$ are parameters depending on the environment and ${\theta _{ik}} = {\rm{ta}}{{\rm{n}}^{ - 1}}\left( {\frac{{{h_k}}}{{{r_{ik}}}}} \right)$ is the elevation angle in radians. Note that the probability of having a NLoS link is represented by ${P_{{\rm{NLoS}}}}{\rm{ = }}1 - {P_{{\rm{LoS}}}}$.\\
\indent In practice, path loss may be affected by specific topology, ground morphology, shadowing effect as well as rapid fluctuations over short time durations. However, following \cite{6863654,7918510,8038014}, we consider the mean effect of path loss instead of its small-scale fluctuations in this paper. 
\begin{figure}
\centering
\includegraphics[width=0.9\linewidth]{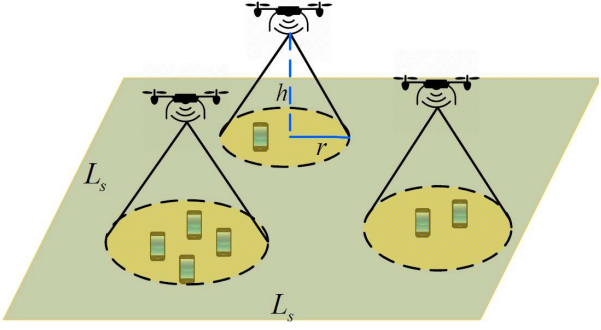}
\caption{System model}
\label{1}
\end{figure}
As indicated in \cite{6863654}, the mean path loss can be modeled as free space path loss plus an excessive path loss which depends on the propagation group. Therefore, the path loss model for LoS and NLoS links in dB are respectively shown as
\begin{eqnarray}
{\rm{P}}{{\rm{L}}_{{\rm{LoS}}}}&=& 20\log \left( {\frac{{4\pi {f_c}{d_{ik}}}}{c}} \right) + {\eta _{{\rm{LoS}}}}\nonumber\\
{\rm{P}}{{\rm{L}}_{{\rm{NLoS}}}}&=& 20\log \left( {\frac{{4\pi {f_c}{d_{ik}}}}{c}} \right) + {\eta _{{\rm{NLoS}}}}
\end{eqnarray}
where ${f_c}$ is the carrier frequency of the system, $c$ is the light speed, ${\eta _{{\rm{LoS}}}}$ and ${\eta _{{\rm{NLoS}}}}$ are the excessive loss for LoS and NLoS links respectively. In addition, ${{d_{ik}}}$ represents the Euclidean distance between user $i$ and the $k$-th aerial BS, which is given by ${d_{ik}}{\rm{ = }}\sqrt {{r_{ik}}^2 + {h_k}^2}$. The weighted path loss between user $i$ and the $k$-th aerial BS, which is a function of both ${{h_k}}$ and ${r_{ik}}$ is then expressed as
\begin{equation}
    {\rm{PL}}\left( {{h_k},{r_{ik}}} \right){\rm{ = }}{P_{{\rm{LoS}}}}{\rm{P}}{{\rm{L}}_{{\rm{LoS}}}} + {P_{{\rm{NLoS}}}}{\rm{P}}{{\rm{L}}_{{\rm{NLoS}}}}
\end{equation}
By substituting (1) and (2) into (3), the above equation yields 
\begin{eqnarray}
{\rm{PL}}\left( {{h_k},{r_{ik}}} \right) = \frac{A}{{1 + a{\rm{ exp}}( - b(\frac{{180}}{\pi }{\rm{tan(}}\frac{{{h_k}}}{{{r_{ik}}}}{\rm{)}} - a))}} + \nonumber\\10 {\rm{log}}\left( {{h_k}^2 + {r_{ik}}^2} \right) + B
\end{eqnarray}
where $A = {\eta _{{\rm{LoS}}}} - {\eta _{{\rm{NLoS}}}}$ and $B = 20{\rm{log}}(\frac{{4\pi }}{c}) + 20\log ({f_c}) + {\eta _{{\rm{NLoS}}}}$. Alternatively, equation (4) can be written as 
\begin{eqnarray}
    {\rm{PL}}\left( {{{\theta _{ik}}},{r_{ik}}} \right) = \frac{A}{{1 + a{\rm{ exp}}( - b(\frac{{180}}{\pi }{\theta _{ik}} - a))}} + \nonumber\\ 20{\rm{log(}}\frac{{{r_{ik}}}}{{\cos ({\theta _{ik}})}}{\rm{)}} + B
\end{eqnarray}
In this paper, the service threshold is defined in terms of the received power of user $i$, which is denoted by $P_r^i$. If the transmit power of the $k$-th aerial BS is $P_t^k$, we have
\begin{equation}
    {P_r^i} = {P_t^k} - {\rm{PL}}\left( {{h_k},{r_{ik}}} \right)
\end{equation}
When the received power of user $i$ with regard to the $k$-th aerial BS is larger than or equal to the threshold value ${P_{\min }}$, user $i$ is covered by the $k$-th aerial BS. When transmit power $P_t^k$ is given, this is equivalent to saying that, user $i$ is covered by the $k$-th aerial BS when the propagation path loss is smaller than or equal to a threshold value ${\gamma _{th}}$. Since the mean effect of path loss is considered in this paper, the coverage area of each aerial BS is a circle area with radius $R_k$, where ${R_k} = {r_{ik}}{|_{{\rm{PL}}\left( {{h_k},{r_{ik}}} \right) = {\gamma _{th}}}}$. It has been shown in \cite{7918510} that, the maximum coverage radius is always associated with a fixed elevation angle ${\theta _{{\rm{opt}}}}$, which only depends on the environment, and is given by
\begin{equation}
    {\theta _{{\rm{opt}}}} = {\tan ^{ - 1}}(\frac{{h_k^*}}{{R_k^*}})
\end{equation}
where ${R_k^*}$ is the maximum coverage radius we can obtain and ${h_k^*}$ is the corresponding optimal altitude of the $k$-th aerial BS that gives ${R_k^*}$. Based on (5) and (7), the threshold path loss can be written as 
\begin{eqnarray}
    {{\gamma _{th}}} = \frac{A}{{1 + a{\rm{ exp}}( - b(\frac{{180}}{\pi }{\theta _{opt}} - a))}} + \nonumber\\20\log (\frac{{R_k^*}}{{\cos ({\theta _{opt}})}}) + B
\end{eqnarray}
from which we can solve ${R_k^*}$ when a specific ${\gamma _{th}}$ is given. In this paper, we consider urban environment whose optimal elevation angle ${\theta _{{\rm{opt}}}}$ equals ${42.44^ \circ }$ \cite{7918510}. Note that when multiple UAVs are deployed, the interference effect needs to be addressed. It is clear that ICI can be intrinsically avoided when there is no overlap between coverage areas of aerial BSs.

\subsection{User Distribution}
Without loss of generality, in this paper, we assume a random user distribution which is obtained through a spatial point process (SPP). We assume three types of SPP, namely homogeneous Poisson process, inhomogeneous Poisson process and Poisson cluster process \cite{illian2008statistical,b2d0c2b554ce4a27a713bdd8770395a0}. The three SPP models are able to describe a majority of user distributions in real scenarios. Let $D$ denote a bounded set, $X(D)$ denote a counting measure of $D$ which calculates the random number of points in $D$, and $\mu (D)$ is a mean measure of $D$, giving the expected number of points. 
\subsubsection{Homogeneous Poisson process}
In homogeneous Poisson process, all user points are uniformly and independently distributed within the target area $W$. The point density equals to a constant $\lambda _s$, which describes the average number of user points generated in a unit area. Therefore, for any user $({x_i},{y_i})$, we have 
\begin{equation}
    P(({x_i},{y_i}) \in S) = \frac{S}{W}
\end{equation}
for any subarea $S$ of the target area $W$. Note that the number of user points generated follows Poisson distribution with  $\mu (D) = {\lambda _s} \cdot W$, which is $X(D) \sim {\rm{Poisson(}}{\lambda _s} \cdot W{\rm{)}}$.
\subsubsection{Inhomogeneous Poisson process}
Inhomogeneous Poisson process is a more general SPP model which introduces inhomogeneity. The constant point density $\lambda _s$ is replaced by an intensity function $\lambda (x,y)$, which varies with locations in the target area. An example intensity function can be 
\begin{equation}
    \lambda (x,y) = c({x^2} + {y^2}) 
\end{equation}
where $c$ is a constant. Then we have
\begin{equation}
    \mu (D) = E\left\{ {X(D)} \right\} = \int_D^{} {\lambda (x,y)} dxdy
\end{equation}
where $E\left\{ . \right\}$ is the expectation operator. The corresponding number of generated user points is thus $X(D) \sim {\rm{Poisson(}}\mu (D){\rm{)}}$, with $\mu (D)$ obtained from (11).
\subsubsection{Poisson cluster process}
Users often gather around points of interest in real scenarios, in which case their distributions involves clustering. In order to describe this kind of user distribution, a Poisson cluster SPP is utilized \cite{b2d0c2b554ce4a27a713bdd8770395a0}. Firstly, a set of 'parent' points ${S_p}$ is generated following homogeneous Poisson process with constant point density ${\lambda _p}$. Then for each $c \in {S_p}$, 'children' points are independently generated following Poisson process with intensity function ${\lambda _c}(x,y)$. Note that 'children' points are distributed in circles around corresponding 'parent' points to form clusters.

\section{proposed successive deployment method with geometrical relaxation (SD-GR)}
In this section, we propose a method based on successive circle placement to find the optimal locations of aerial BSs such that a maximum user coverage probability can be achieved. Following \cite{7486987,7762053}, and as shown in Fig. 2(a), we assume that all UAVs have the same coverage radius $R$ and are thus deployed in the same altitude $H$ when a specific path loss requirement is given, that is 
\begin{eqnarray}
   &&{h_k} = H, k = 1,2,...,K\\
   &&{R_k} = R, k = 1,2,...,K\nonumber\\
   &&H = R\tan ({\theta _{opt}})\nonumber
\end{eqnarray}
with $R$ calculated from (8). Therefore, placing multiple UAVs is equivalent to placing multiple circles in the horizontal plane such that the number of enclosed user points is maximized. UAVs are placed in a successive method, where at each step the placement of the aerial BS aims to cover the maximum number of remaining users in the target area while ensuring that there is no overlap in coverage areas with all previously deployed BSs. The location of the first UAV can be found by utilizing the method proposed in \cite{7918510}. Let ${C_1}$ denote the coverage area of the first UAV and Boolean variable ${u_i} \in \{ 0,1\} $, $i \in \cal{M}$ denotes the status of user $i$ such that the user is enclosed by ${C_1}$ when ${u_i} = 1$ and is not covered by the first UAV when ${u_i} = 0$. Then the circle placement problem is formulated as
\begin{eqnarray}
    &&\mathop {{\rm{maximize}}}\limits_{{x_{c1}},{y_{c1}},{u_i}} \sum\limits_{i \in \cal{M}} {{u_i}}\\
    &&{\rm{subject}}{\kern 1pt}{\kern 1pt} {\rm{to}}\nonumber\\
    &&{{\rm{(}}{x_i} - {x_{c1}}{\rm{)}}^2} + {({y_i} - {y_{c1}})^2} \le {R^2} + M(1 - {u_i}),\forall i \in \cal{M}\nonumber\\
    &&{u_i} \in \{ 0,1\} ,\forall i \in \cal{M}\nonumber
\end{eqnarray}
where $({x_{c1}},{y_{c1}})$ is the horizontal location of the first UAV, i.e. the center of the coverage region, and $M$ is a large constant that can be any value larger than the square of the largest distance between any two points in the target area. It can be seen that the first constraint of (13) reduces to
\begin{equation}
    {{\rm{(}}{x_i} - {x_{c1}}{\rm{)}}^2} + {({y_i} - {y_{c1}})^2} \le {R^2},\forall i \in \cal{M}
\end{equation}
when ${u_i} = 1$ which is equivalent to saying that user $i$ is covered by the first UAV, and the objective function of (13) is increased by 1 correspondingly. 
\begin{figure}
\centering
\subfigure[non-convex region]{
\label{convex a}
\includegraphics[width=1.4in]{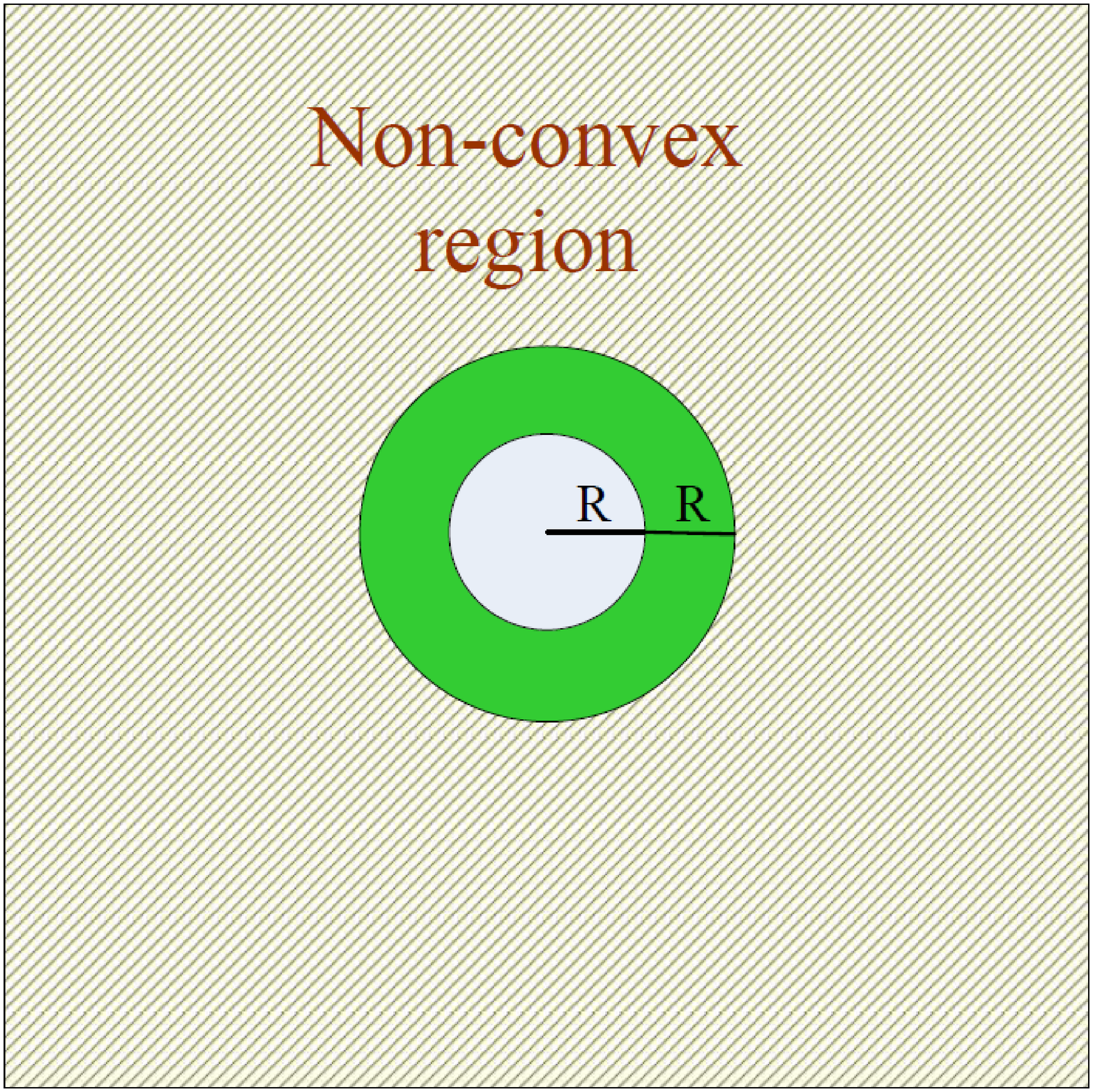}
}
\hspace{0.2in}
\subfigure[four convex regions]{
\label{convex b}
\includegraphics[width=1.4in]{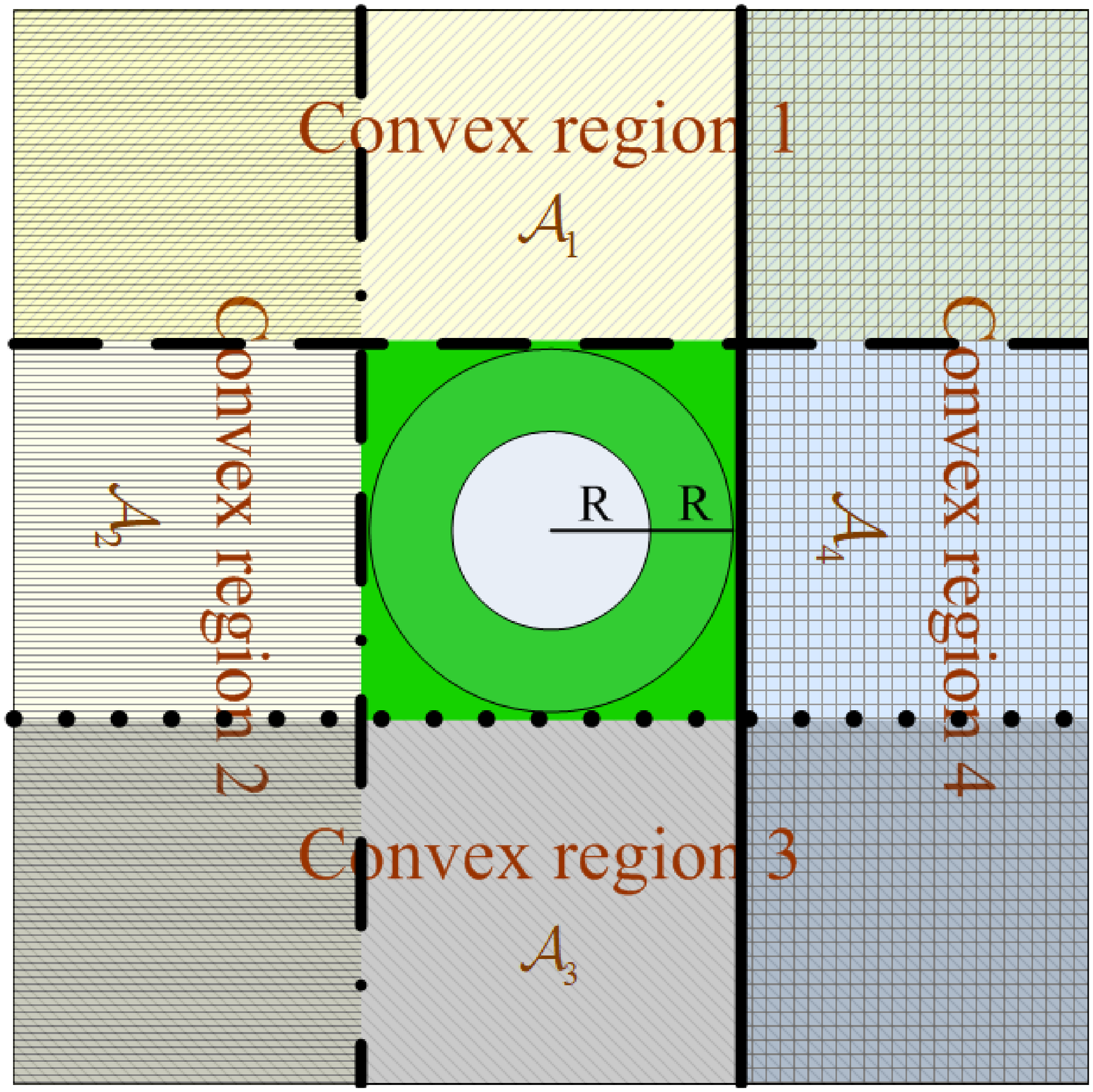}
}
\caption{Converting the non-convex region into convex regions with geometrical relaxation}
\label{geomoetrical explanation}
\end{figure}
In addition, the inequality of the constraint is still satisfied when ${u_i} = 0$ due to the very large constant $M$ \cite{7918510}.\\
\indent However, when we want to deploy the second UAV, an additional constraint ensuring no overlapping between coverage areas, and therefore no ICI, is needed. To ensure this, the distance between the two UAVs in the horizontal dimension should be larger than 2$R$. Therefore, the placement of the second UAV is formulated as
\begin{eqnarray}
    &&\mathop {{\rm{maximize}}}\limits_{{x_{c2}},{y_{c2}},{u_i}} \sum\limits_{i \in \cal{M}} {{u_i}}\\
    &&{\rm{subject}}{\kern 1pt}{\kern 1pt} {\rm{to}}\nonumber\\
    &&{{\rm{(}}{x_i} - {x_{c2}}{\rm{)}}^2} + {({y_i} - {y_{c2}})^2} \le {R^2} + M(1 - {u_i}),\forall i \in \cal{M}\nonumber\\
    &&{{\rm{(}}{x_{c2}} - {x_{c1}}{\rm{)}}^2} + {({y_{c2}} - {y_{c1}})^2} \ge 4{R^2}\nonumber\\
    &&{u_i} \in \{ 0,1\} ,\forall i \in \cal{M}\nonumber
\end{eqnarray}
where $({x_{c2}},{y_{c2}})$ is the horizontal location of the second UAV. Unfortunately, the additional constraint is non-convex which makes (15) extremely hard to solve. Although the integer variables in (13) can be addressed with advanced mixed integer programming techniques, using solvers such as MOSEK \cite{7918510}, the optimization problem (15) which is a MINLP problem with non-convex constraint can not be straightforwardly solved. Even if we apply semidefinite relaxation (SDR) techniques to convert the quadratic programs into the form of semidefinite matrix which makes the non-convex constraint of (15) convex, a problem with both positive semidefinite matrix and integer variables is still unsolvable with existing techniques \cite{5447068}.\\
\indent In Fig. 2(a), the circle in white with radius $R$ represents the coverage area of the first aerial BS, and the circle with radius $2R$ represents the area where there cannot be any placement of additional UAVs without inflicting ICI. Accordingly, the region outside the green circle with radius $2R$ is the geometrical representation of the non-convex constraint in (15). We notice that such a non-convex region which specifies all the feasible locations of the second UAV in the horizontal dimension can be divided into four linear regions which are convex. This is done by approximating the green circular area by a square area such that the original green circle is surrounded by the square as illustrated in Fig. 2(b). Therefore, instead of solving (15), we can solve four MINLP problems with different linear constraints, and each of the four problems has the following form
\begin{eqnarray}
    &&\mathop {{\rm{maximize}}}\limits_{{x_{c2}},{y_{c2}},{u_i}} \sum\limits_{i \in \cal{M}} {{u_i}}\\
    &&{\rm{subject}}{\kern 1pt}{\kern 1pt} {\rm{to}}\nonumber\\
    &&{{\rm{(}}{x_i} - {x_{c2}}{\rm{)}}^2} + {({y_i} - {y_{c2}})^2} \le {R^2} + M(1 - {u_i}),\forall i \in \cal{M}\nonumber\\
    &&{y_{c2}} \ge {y_{c1}} + 2R,{\rm{if }}\left( {{x_{c2}},{y_{c2}}} \right) \in \mathcal{A}_1\nonumber\\
    &&{x_{c2}} \le {x_{c1}} - 2R,{\rm{if }}\left( {{x_{c2}},{y_{c2}}} \right) \in \mathcal{A}_2\nonumber\\
    &&{y_{c2}} \le {y_{c1}} - 2R,{\rm{if }}\left( {{x_{c2}},{y_{c2}}} \right) \in \mathcal{A}_3\nonumber\\
    &&{x_{c2}} \ge {x_{c1}} + 2R,{\rm{if }}\left( {{x_{c2}},{y_{c2}}} \right) \in \mathcal{A}_4\nonumber\\
    &&{u_i} \in \{ 0,1\} ,\forall i \in \cal{M}\nonumber
\end{eqnarray}
where $\mathcal{A}_1$, $\mathcal{A}_2$, $\mathcal{A}_3$ and $\mathcal{A}_4$ are the four convex regions shown in Fig. 2(b). The maximum number of covered users as well as the location of the second UAV is then found among the results of four MINLP problems. Note that the overlapping in the four convex regions will not affect the final result and is thus allowed. If the optimal location of the second UAV is inside the overlapping area, two of the four optimization problems will give the same solution which contains a maximum number of covered users. However, the effective area for placing the second UAV is slightly reduced as a result of considering the infeasible region as a square region.\\
\begin{figure}
\centering
\includegraphics[width=1\linewidth]{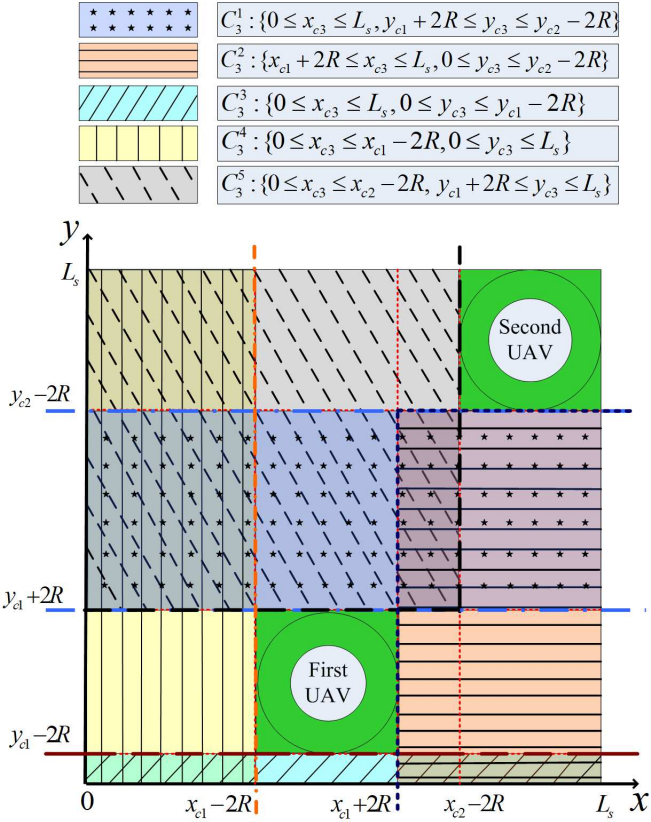}
\caption{An example of feasible region definition, with two deployed aerial BSs, for the positioning of the third BS}
\label{3}
\end{figure}

\indent The optimization problem of placing the $k$-th UAV ($k > 1$) is formulated as 
\begin{eqnarray}
    &&\mathop {{\rm{maximize}}}\limits_{{x_{ck}},{y_{ck}},{u_i}} \sum\limits_{i \in \cal{M}} {{u_i}}\\
    &&{\rm{subject}}{\kern 1pt}{\kern 1pt} {\rm{to}}\nonumber\\
    &&{{\rm{(}}{x_i} - {x_{ck}}{\rm{)}}^2} + {({y_i} - {y_{ck}})^2} \le {R^2} + M(1 - {u_i}),\forall i \in \cal{M}\nonumber\\
    &&{({x_{ck}} - {x_{cj}})^2} + {({y_{ck}} - {y_{cj}})^2} \ge 4{R^2},{\rm{ }}j = 1,2,...,k - 1\nonumber\\
    &&{u_i} \in \{ 0,1\} ,\forall i \in \cal{M}\nonumber
\end{eqnarray}
where $({x_{ck}},{y_{ck}})$ and $({x_{cj}},{y_{cj}})$ denote the horizontal location of the $k$-th UAV and the $j$-th UAV respectively. For each of the $k-1$ non-convex constraints, geometrical relaxation is utilized to convert it into four linear constraints as illustrated above. As each linear constraint only specifies one of the four feasible regions generated by a certain UAV, the true feasible region for the $k$-th UAV is the intersection of $(k - 1)$ sets, where each set is the union of four feasible regions. Evidently, the total feasible region which is non-convex consists of several convex regions of rectangular shape as shown in Fig. 3. 
\begin{algorithm}[t!]
\caption{Algorithm for placing the $k$-th UAV with geometrical relaxation}
\begin{algorithmic}[1]
\renewcommand{\algorithmicrequire}{\textbf{Inputs:}}
\REQUIRE
user locations, $({x_i},{y_i}) \in \cal{M}$; radius of coverage area, $R$; locations of all deployed UAVs $({x_{cj}},{y_{cj}}), {\rm{ }}j = 1,2,...,k-1$
\renewcommand{\algorithmicensure}{\textbf{Output:}}
\ENSURE
 number of users covered by the $k$-th UAV, ${U_k}$; the location of the $k$-th UAV, $({x_{ck}},{y_{ck}})$
\renewcommand{\algorithmicrequire}{\textbf{Initialization:}} 
\REQUIRE
$j$=1;
$z$=1;
$m$=0.
\WHILE{$j < k$}
\STATE
converting the constraint ${({x_{ck}} - {x_{cj}})^2} + {({y_{ck}} - {y_{cj}})^2} \ge 4{R^2}$ into four linear constraints which are ${x_{ck}} \ge {x_{cj}} + 2R$, ${x_{ck}} \le {x_{cj}} - 2R$, ${y_{ck}} \ge {y_{cj}} + 2R$, and ${y_{ck}} \le {y_{cj}} - 2R$ respectively.
\STATE
$j = j + 1$.
\ENDWHILE
\STATE
For each of the $k-1$ UAVs, one of the four linear constraints is selected to form the intersection of these $k-1$ regions.  A total of ${4^{k - 1}}$ intersections are generated and denoted as ${C_z},z = 1,2,...,{4^{k - 1}}$.
\WHILE{$z < {4^{k - 1}}$}
\IF{${C_z} = \emptyset$}
\STATE
eliminate ${C_z}$
\ELSIF{${C_z} \subseteq {C_q},q = 1,2,...,{4^{k - 1}},q \ne z$}
\STATE
eliminate ${C_z}$
\ELSE
\STATE
$m = m + 1$.
\STATE
$C_k^m = {C_z}$.
\STATE
obtain $(x_{ck}^m,y_{ck}^m)$, and ${N_m}$ by solving (18)
\ENDIF
\STATE
$z = z + 1$.
\ENDWHILE
\STATE
${N_{\max }} = \max ({N_m}),{ m = 1,2,}...{\rm{,}}N_M^k$.
\STATE
${U_k}$=${N_{\max }}$.
\STATE
$({x_{ck}},{y_{ck}}) = (x_{ck}^m,y_{ck}^m){|_{{N_m} = {N_{\max }}}}$.
\end{algorithmic}
\end{algorithm}
The total number of convex regions depends on specific deployment but can be found through an elimination method. To be specific, for each of the $k-1$ UAVs, one of the four generated feasible regions is selected and logic function is used to find the intersection of these $k-1$ selected regions to make sure the coverage area of the next aerial BS does not interfere with any previously deployed aerial BSs. A total of ${{\rm{4}}^{k - 1}}$ intersections should be generated and we denote each intersection as ${C_z},z = 1,2,...,{4^{k - 1}}$. After obtaining all the ${{\rm{4}}^{k - 1}}$ intersections, we eliminate all sets which are null sets, i.e. ${C_z} = \emptyset$ or sets which turns out to be subsets of other generated sets, i.e. ${C_z} \subseteq {C_q},q = 1,2,...,{4^{k - 1}},q \ne z$, and the remaining intersections are the feasible regions we should search for. An example is shown in Fig. 3. Here, 5 feasible regions are found among 16 intersections for deploying the third aerial BS after eliminating all null sets and subsets. We denote the total number of feasible regions for deploying the $k$-th UAV as ${N_M^k}$. In this case, (17) can be reformulated as ${N_M^k}$ MINLP problems, each has the following form
\begin{eqnarray}
    &&\mathop {{\rm{maximize}}}\limits_{{x_{ck}^m},{y_{ck}^m},{u_i}} \sum\limits_{i \in \cal{M}} {{u_i}}\\
    &&{\rm{subject}}{\kern 1pt}{\kern 1pt} {\rm{to}}\nonumber\\
    &&{{\rm{(}}{x_i} - {x_{ck}^m}{\rm{)}}^2} + {({y_i} - {y_{ck}^m})^2} \le {R^2} + M(1 - {u_i}),\forall i \in \cal{M}\nonumber\\
    &&({x_{ck}^m},{y_{ck}^m}) \in C_k^m\nonumber\\
    &&{u_i} \in \{ 0,1\} ,\forall i \in \cal{M}\nonumber
\end{eqnarray}
where $C_k^m$ is the $m$-th feasible region of the $k$-th aerial BS, $(x_{ck}^m,y_{ck}^m)$ is the optimal location of the $k$-th UAV in region $C_k^m$, $m = 1,2,...,{N_M^k}$. If we denote the number of covered users by solving the $m$-th optimization problem as ${N_m}$, and denote the maximum ${N_m}$ for all $m$ as ${N_{\max }}$, we have $({x_{ck}},{y_{ck}}) = (x_{ck}^m,y_{ck}^m){|_{{N_m} = {N_{\max }}}}$. In addition, ${U_k} = {N_{\max }}$, where ${U_k}$ denotes the number of covered users by the $k$-th aerial BS. For clarity, the proposed geometrical relaxation method is summarized in Algorithm 1.

\section{Proposed simultaneous deployment method with k-means clustering (SD-KM)}
The shortcoming of the above proposed algorithm is that it introduces exponentially increasing computational complexity due to the need to solve ${{\rm{4}}^{k - 1}}$ logic combination operations as well as multiple MINLP problems for finding the optimal location of the $k$-th UAV, which makes its use prohibitively complex when a large number of aerial BSs are needed. As a result, there is a strong motivation for a mechanism in which multiple aerial BSs can be deployed simultaneously without introducing non-convex constraints. In this section, a method which simultaneously deploys multiple UAVs is proposed with the help of $K$-means clustering.\\
\indent $K$-means clustering, which is a well-known partitional clustering method has been utilized in a variety of disciplines \cite{Xu:2009:CLU:1483087}. In our particular scenario, we notice that the whole target coverage area can be divided into $K$ subareas with boundaries forming the Voronoi diagram, by assigning user points into $K$ clusters. The intelligent division of the target area brings great benefit to the deployment of multiple UAVs in several senses. First, each subarea which is bounded by few straight lines or line segments is a convex region. Within each convex region, we can solve an optimization problem similarly to (13) to find the best location of a UAV so a maximum number of user points within that subarea is enclosed. In addition, the boundary lines of Voronoi diagram ensure that the circles placed in each subarea will not overlap with each other, so the ICI is intrinsically avoided. Furthermore, a number of $K$ UAVs can be deployed simultaneously in corresponding subareas, so the latency and dependence on previously deployed aerial BSs with successive circle placement is solved. Last but not the least, $K$-means clustering finds potential clustering properties among user points. The clustering properties indicate a reasonable number of UAVs to be deployed, hence avoiding the use of excessive UAVs and saving both costs and power. The details of the proposed method are illustrated in the following two subsections.

\subsection{Applying $K$-means clustering and partitioning the target area}
We assume the user set $\cal{M}$ contains a total of ${U_{{\rm{tot}}}}$ users and we denote arrays storing the location of user points by ${{\bf{u}}_i}$, where ${{\bf{u}}_i} = [{x_i},{y_i}]$, $i = 1,2,...,{U_{{\rm{tot}}}}$. The aim of applying $K$-means clustering is organizing the ${U_{{\rm{tot}}}}$ user points into $K$ clusters $\mathcal{C}_k$, $k \in [1,K]$, with the $k$-th cluster, $k = 1,2,...,K$, containing ${N_k}$ user points, out of which ${U_k} \le {N_k}$ user points are covered. The partition is very much based on a sum-of-squared-error criterion \cite{Xu:2009:CLU:1483087} which is defined as
\begin{equation}
    e = \sum\limits_{k = 1}^K {\sum\limits_{i = 1}^{{U_{{\rm{tot}}}}} {{u_{ki}}} } {\left\| {{{\bf{u}}_i} - {{\bf{m}}_k}} \right\|^2}
\end{equation}
where $\left\| . \right\|$ denotes the frobenius norm of a vector. ${{u_{ki}}}$ here is a Boolean variable, indicating the state of $i$-th user of the $k$-th cluster. We have ${u_{ki}} = 1$ when $({x_i},{y_i}) \in \mathcal{C}_k$ and ${u_{ki}} = 0$ otherwise. In addition, ${{\bf{m}}_k}$ is an array storing the center location of the $k$-th cluster. Note that the center of a certain cluster is obtained by calculating the mean value of all user points classified into that cluster, which can be written as
\begin{equation}
    {{\bf{m}}_k} = [{m_{kx}},{m_{ky}}]  = \left\{ {\frac{1}{{{N_k}}}\sum\limits_{i = 1}^{{U_{{\rm{tot}}}}} {{u_{ki}}{x_i}} ,\frac{1}{{{N_k}}}\sum\limits_{i = 1}^{{U_{{\rm{tot}}}}} {{u_{ki}}{y_i}} } \right\}
\end{equation}
where $k = 1,2,...,K$. Then the procedure of applying the $K$-means clustering is concluded in four steps.
\begin{enumerate}
    \item  Randomly choose $K$ points in the target area as the center points and store them in ${{\bf{m}}_k}$, $k = 1,2,...,K$
    \item Allocate each user point in $\cal{M}$ to the cluster with the closest center $\mathcal{C}_j$, i.e., $({x_i},{y_i}) \in \mathcal{C}_j$ when the Euclidean distance between ${{\bf{u}}_i}$ and the center of cluster $j$ is smaller than the Euclidean distance between ${{\bf{u}}_i}$ and any other cluster centers.
\begin{eqnarray}
    &&\left\| {{{\bf{u}}_i} - {{\bf{m}}_j}} \right\| < \left\| {{{\bf{u}}_i} - {{\bf{m}}_k}} \right\|, \\&&i = 1,2,...,{U_{{\rm{tot}}}},{\kern 1pt} {\kern 1pt} k = 1,2,...,K, {\kern 1pt} {\kern 1pt}j \ne k\nonumber
\end{eqnarray}
    \item Recalculate the cluster centers as the mean position of all user points in each cluster.  
    \item Repeat the above two steps until there is no change for ${{\bf{m}}_k}$.
\end{enumerate}
 
\indent Note that $K$-means clustering assumes the number of clusters to generate is known, which is not true for our case. Either excessive or inadequate number of generated subareas can affect the number of users covered. The number of required clusters highly depends on user distributions, so variable $K$ might be utilized for various scenarios. For all cases, we first start with a maximum $K$ value, denoted as ${K_{\max }}$, making sure adequate number of subareas are generated even for the case showing the least clustering property (uniform distribution). Then we try to reduce the number of $K$ since user points might gather at some locations. As excessive partition can split a single cluster into several parts, which severely deteriorate the performance of the proposed method, we set a threshold ${d_{th}}$ indicating the minimum allowed distance between two cluster centers. If $\min (\left\| {{{\bf{m}}_j} - {{\bf{m}}_k}} \right\|) < {d_{th}}$ , $j \ne k$, this signifies that some of the generated clusters are too small as a result of using too large $K$ value. Then we reduce the value of $K$ by one and reapply the $K$-means clustering. The above procedure continues until we have $\min (\left\| {{{\bf{m}}_j} - {{\bf{m}}_k}} \right\|) \ge {d_{th}}$.

\subsection{Solving the optimization problem within each region}
After partitioning the user points into $K$ clusters and, subsequently, dividing the whole target area into $K$ subareas, we first need to find the largest allowed coverage area within each subarea to avoid ICI. As the shape of each subarea is a polygon, it is likely that certain subareas can only accommodate circles with radii smaller than $R$ . Assume the $k$-th subarea is formed with ${S_k}$ line segments or straight lines, each line is expressed in the form of $y = {a_{kl}}x + {b_{kl}}$, $l = 1,2,...,{S_k}$, where ${a_{kl}}$ and ${b_{kl}}$ represents the slope and offset of the $l$-th boundary line of the $k$-th subarea respectively. It is known that for any point $({x_d},{y_d})$, if ${y_d} - {a_{kl}}{x_d} - {b_{kl}} < 0$, the point is in the region below the line. On the contrary, if ${y_d} - {a_{kl}}{x_d} - {b_{kl}} > 0$, the point is in the region above the line. 
We note that the boundary lines of each subarea also implicates a feasible region for placing the circle center as well as restricting the length of radius. To be specific, the distance between the circle center and each boundary line should be no smaller than the length of radius of the circle. Therefore, the region for placing the circle can be found by shifting the boundary lines of each subarea. If the cluster center ${{\bf{m}}_k}$ is in the region below a certain boundary line of the $k$-th subarea, the corresponding new line specifying the region for placing the circle center can be obtained by shifting the line downward along the y-axis by ${L_{kl}}$. Similarly, shifting the original boundary line upward along the y-axis by ${L_{kl}}$ leads to the corresponding new line when ${{\bf{m}}_k}$ is in the region above the original boundary line. Here, ${L_{kl}}$ denotes the length to be shifted along the y-axis of the $l$-th boundary line of the $k$-th subarea, and is calculated through
\begin{equation}
    {L_{kl}} = \frac{{R_{\max }^k}}{{\cos (\left| {{a_{kl}}} \right|)}},{\kern 1pt}{\kern 1pt}k = 1,2,...,K,{\kern 1pt}{\kern 1pt}l = 1,2,...,{S_k}
\end{equation}
where $R_{\max }^k$ denotes the maximum allowed radius of the circle placed in the $k$-th subarea and $\left| . \right|$ calculates the amplitude of a number. Therefore, $R_{\max }^k$ can be found by solving the following optimization problem.
\begin{eqnarray}
    &&\mathop {{\rm{maximize}}}\limits_{{x_{ck}},{y_{ck}},R_{\max }^k} {\rm{ }}R_{\max }^k\\
    &&{\rm{subject}}{\kern 1pt}{\kern 1pt} {\rm{to}}\nonumber\\
    &&{y_{ck}} - {a_{kl}}{x_{ck}} - {b_{kl}} + {L_{kl}} \le 0,\nonumber\\
    &&{\rm{if}}{\kern 1pt}{\kern 1pt}{\kern 1pt}{\kern 1pt}{m_{ky}} - {a_{kl}}{m_{kx}} - {b_{kl}} \le 0 \nonumber\\
    &&{y_{ck}} - {a_{kl}}{x_{ck}} - {b_{kl}} - {L_{kl}} \ge 0, \nonumber\\
    &&{\rm{if}}{\kern 1pt}{\kern 1pt}{\kern 1pt}{\kern 1pt}{m_{ky}} - {a_{kl}}{m_{kx}} - {b_{kl}} \ge 0\nonumber\\
    &&k = 1,2,...,K,{\kern 1pt}{\kern 1pt}{\kern 1pt}{\kern 1pt}l = 1,2,...,{S_k}\nonumber
\end{eqnarray}
The radius of the $k$-th circle is then ${R_k} = \min (R,{R_{\max }^k})$. With fixed radii, we can find the optimal placement of circles for covering maximum number of user points within their corresponding subareas by solving the following problem
\begin{eqnarray}
&&\mathop {{\rm{maximize}}}\limits_{{x_{ck}},{y_{ck}},{u_i}} \sum\limits_{i \in \cal{M}} {{u_i}}\\
&&{\rm{subject}}{\kern 1pt}{\kern 1pt} {\rm{to}}\nonumber\\
&&{{\rm{(}}{x_i} - {x_{ck}}{\rm{)}}^2} + {({y_i} - {y_{ck}})^2} \le {{R_k}^2} + M(1 - {u_i}),\forall i \in \cal{M}\nonumber\\  
&&{y_{ck}} - {a_{kl}}{x_{ck}} - {b_{kl}} + \frac{{{R_k}}}{{\cos (\left| {{a_{kl}}} \right|)}} \le 0, \nonumber\\
&&{\rm{if}}{\kern 1pt}{\kern 1pt}{\kern 1pt}{\kern 1pt}{m_{ky}} - {a_{kl}}{m_{kx}} - {b_{kl}} \le 0\nonumber\\
&&{y_{ck}} - {a_{kl}}{x_{ck}} - {b_{kl}} - \frac{{{R_k}}}{{\cos (\left| {{a_{kl}}} \right|)}} \ge 0, \nonumber\\
&&{\rm{if}}{\kern 1pt}{\kern 1pt}{\kern 1pt}{\kern 1pt}{m_{ky}} - {a_{kl}}{m_{kx}} - {b_{kl}} \ge 0\nonumber\\
&&{u_i} \in \{ 0,1\} ,\forall i \in \cal{M}\nonumber\\
&&k = 1,2,...,K,{\kern 1pt}{\kern 1pt}{\kern 1pt}{\kern 1pt}l = 1,2,...,{S_k}\nonumber
\end{eqnarray}
The above optimization problem is a MINLP problem without non-convex constraints and is nearly as easy to solve as (13). 
\begin{figure}
\centering
\subfigure[]{
\label{flexibility}
\includegraphics[width=1.4in]{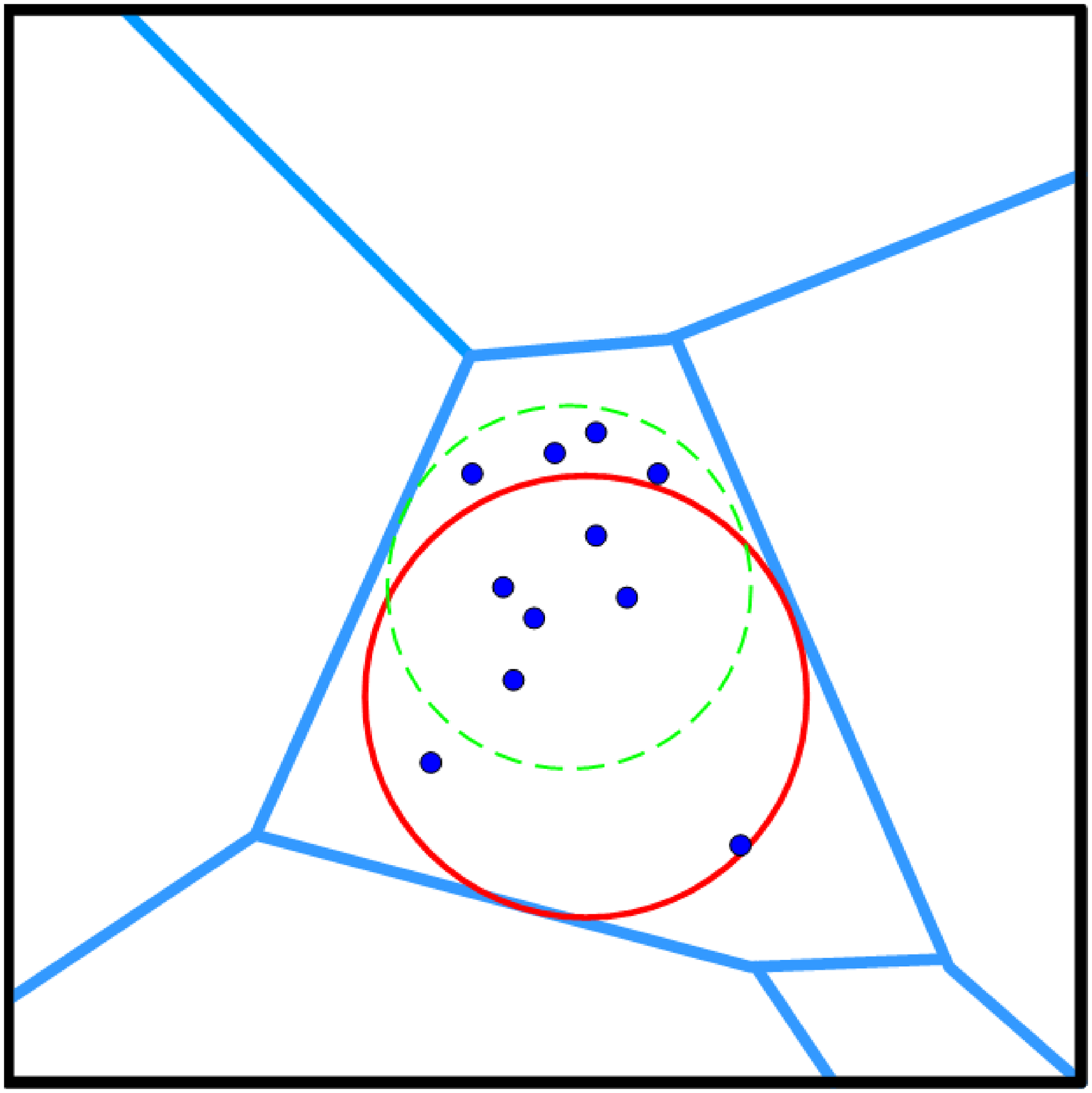}
}
\hspace{0.2in}
\subfigure[]{
\label{less power}
\includegraphics[width=1.4in]{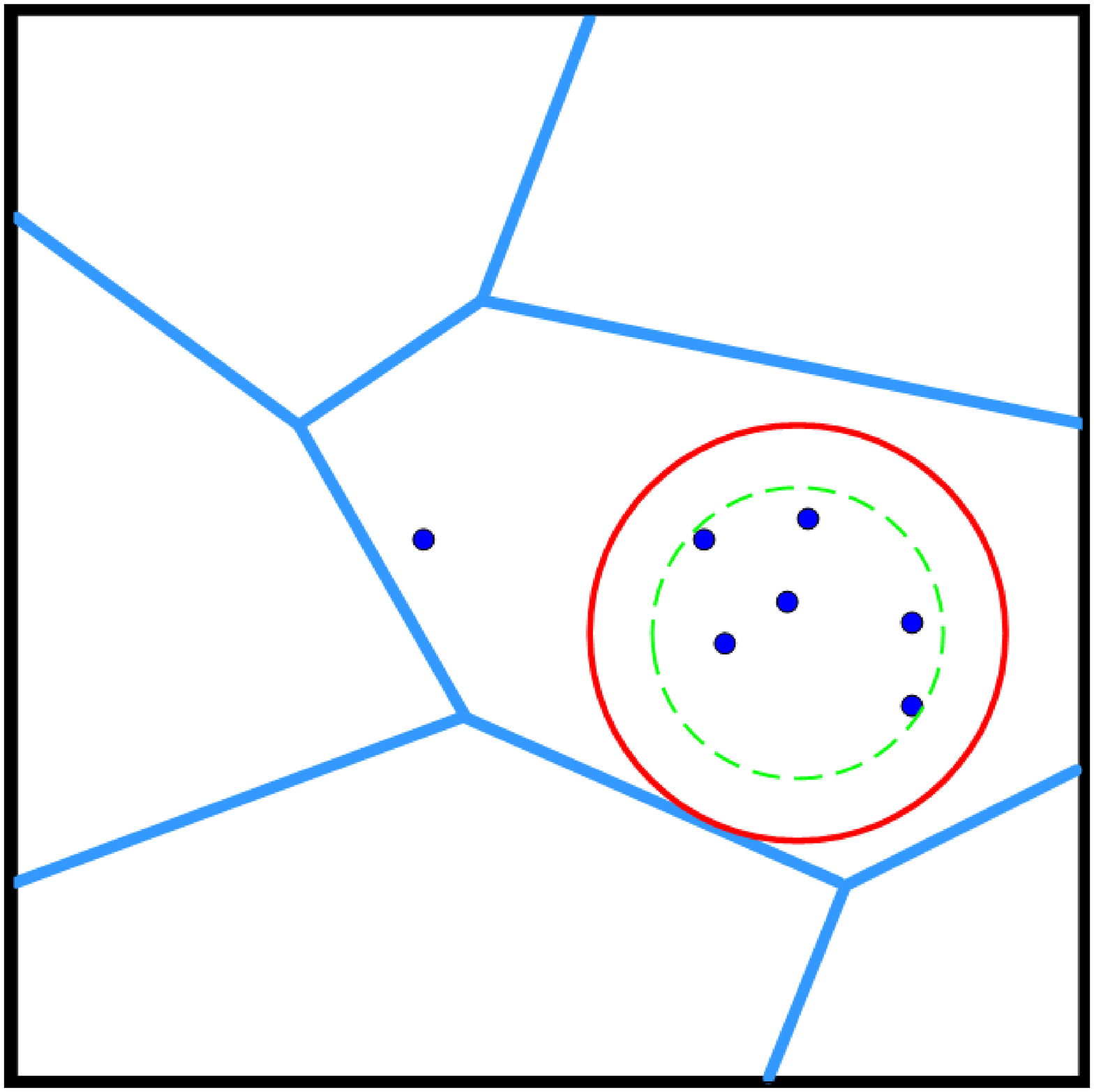}
}
\caption{The case for optimizing the radius in $K$-means circle placement algorithm: (a) flexibility in reaching additional users, (b) reducing power for a given user coverage area.}
\label{improvement}
\end{figure}
Note that with the help of $K$-means clustering, we only need to solve $K$ optimization problems of this type and $K < {K_{\max }}$ is obtained in most cases as illustrated in the first subsection. 

\section{Energy efficient simultaneous deployment method with variable radius (SD-KMVR)}
\begin{algorithm}[t!]
\caption{Iterative algorithm for placing the $k$-th UAV}
\begin{algorithmic}[1]
\renewcommand{\algorithmicrequire}{\textbf{Inputs:}}
\REQUIRE
Initial radius ${R_k}$; an intermediate value storing the change of radius, ${r_{it}}$.
\renewcommand{\algorithmicensure}{\textbf{Output:}}
\ENSURE
 Set containing all covered user points, ${\cal{M}}_{{\mathop{\rm cov}} }^k$; the location of the $k$-th UAV, $({x_{ck}},{y_{ck}})$; the optimal radius $r_k$.
\renewcommand{\algorithmicrequire}{\textbf{Initialization:}} 
\REQUIRE
${r_{it}} = 0$, ${r_k} = {R_k}$
\WHILE{${r_{it}} \ne {r_k}$}
\STATE
${r_{it}} = {r_k}$
\STATE
obtain $({x_{ck}},{y_{ck}})$ and ${\cal{M}}_{{\mathop{\rm cov}} }^k$ by solving (24) and replacing ${R_k}$ with ${r_{it}}$.
\STATE
obtain ${r_k}$ by solving (25).
\ENDWHILE
\end{algorithmic}
\end{algorithm}
In the preceding section, a simultaneous deployment method has been proposed to take advantage of dividing the whole target area into $K$ convex subareas. However, the proposed method based on $K$-means clustering can be further improved in terms of both the coverage probability and power efficiency. First, the maximum circle area does not always cover a maximum number of user points in a irregular polygon region, so variable radius may introduce further improved performance gain. As can be seen in Fig. 4(a), user points may gather in a relative narrow region where circles with large radii can not reach. More user points can thus be enclosed when a smaller coverage area is placed. Furthermore, the radii of coverage areas and hence the transmit power of aerial BSs can be further reduced because there might be no user points right on the border of the circles. In Fig. 4(b), original circle in red obtained by the SK-KM in the previous section, can be shrunk into the circle in green which covers the same set of user points with a reduced transmit power. \\
\begin{figure}
\centering
\includegraphics[width=1\linewidth]{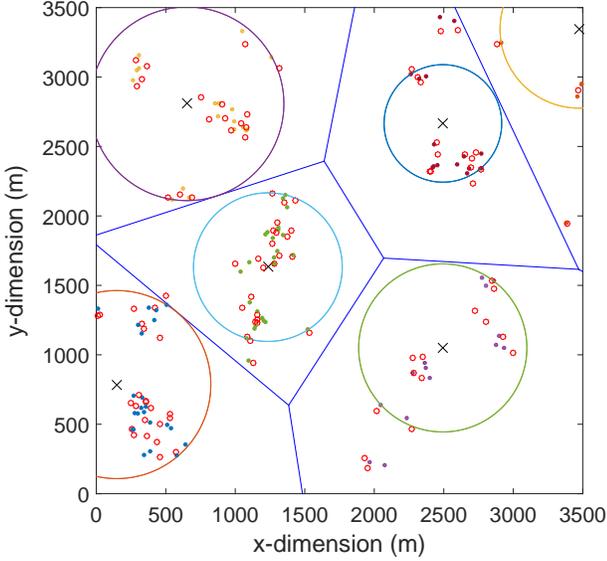}
\caption{SD-KM with inaccurate ULI}
\label{7}
\end{figure}
\indent In order to address both problems at the same time, an iterative algorithm is proposed. We assume aerial BS has a minimum allowed coverage area with radius ${R_{\min }}$, then the variable radius of the $k$-th circle ${r_k}$ has a range of ${R_{\min }} \le {r_k} \le {R_k}$. We first obtain the circle center $({x_{ck}},{y_{ck}})$ as well as the set containing the covered user points, denoted as ${\cal{M}}_{{\mathop{\rm cov}} }^k$ with size ${U_k}$ by solving (24) with radius ${R_k}$. Then, we fix both $({x_{ck}},{y_{ck}})$ and ${\cal{M}}_{{\mathop{\rm cov}} }^k$, aiming to find the minimum ${r_k}$ which can cover the same set of user points by solving the problem,
\begin{eqnarray}
&&{\rm{minimize }}{\kern 1pt}{\kern 1pt}{\kern 1pt}{\kern 1pt}{r_k}\\
&&{\rm{subject}}{\kern 1pt}{\kern 1pt} {\rm{to}}\nonumber\\
&&{r_k}^2 \ge {({x_i} - {x_{ck}})^2} + {({y_i} - {y_{ck}})^2},\forall i \in {\cal{M}}_{{\mathop{\rm cov}} }^k\nonumber\\
&&{R_{\min }} \le {r_k} \le {R_k}\nonumber
\end{eqnarray}
After solving ${r_k}$, we replace ${R_k}$ with ${r_k}$ and solve (24) again to find the updated user points enclosed by the new circle. The above procedure repeats until the radius ${r_k}$ does not change anymore. Algorithm 2 summaries the proposed iterative algorithm. In addition, as the radii of the $K$ coverage areas are reduced and all the aerial BSs are always placed at a position with fixed elevation angle ${\theta _{opt}}$, the altitude of the $k$-th UAV can be found by
\begin{equation}
    {h_k} = {r_k}\tan ({\theta _{opt}})
\end{equation}
Therefore, the 3-D location of the $k$-th aerial BS is marked as $({x_{ck}},{y_{ck}},{h_k})$. Furthermore, the reduced radii also reduce the path loss between UAVs and users according to (8) and thus reduce the required transmit power of aerial BSs since we have
\begin{equation}
    P_t^k = {P_{{\rm{min}}}} + {\rm{PL}}({r_k})
\end{equation}
where $P_t^k$ is the required transmit power of the $k$-th aerial BS and ${P_{\min }}$ is the threshold receive power, below which the communication link is failed. The total required power of the system is thus the sum of transmit power of all aerial BSs which is a function of both $K$ and ${r_k}$,
\begin{equation}
    {P_{total}} = \sum\limits_{k = 1}^K {P_t^k}  = K{P_{\min }} + \sum\limits_{k = 1}^K {\rm{PL}}({r_k})
\end{equation}
Clearly, the iterative algorithm is more power efficient than the proposed SD-KM algorithm at the cost of increasing the computational complexity.

\section{Imperfect ULI and Robust Deployment}
In the practical UAV deployment the ULI may contain errors. Consequently, the coverage probability of the proposed techniques may decrease drastically as a result of estimating user locations inaccurately. In this section, we formulate a robust technique which is applicable to both SD-KM and SD-KMVR to preserve the best coverage performance in the existence of inaccurate ULI. We model the estimated location of user $i$ as $(\mathop {{x_i}}\limits^ \sim  ,\mathop {{y_i}}\limits^ \sim  ) = ({x_i} + {e_{xi}},{y_i} + {e_{yi}})$, where ${e_{xi}}$ and ${e_{yi}}$ are estimation errors following Gaussian distribution with zero mean and standard deviation $\sigma$ in meters. For illustration, Fig. 5 shows an example deployment of SD-KMVR in the existence of imperfect ULI with $\sigma  = 50$ m and ${L_s} = 3500$ m. The small circles in red represent real user locations, and SD-KMVR is applied based on estimated user locations represented by dots.
\begin{algorithm}[t!]
\caption{Robust deployment of aerial BSs}
\begin{algorithmic}[1]
\renewcommand{\algorithmicrequire}{\textbf{Inputs:}}
\REQUIRE
Placement details obtained from SD-KM or SD-KMVR technique: radius of coverage areas, ${R_k}$; horizontal location of aerial BSs, $({x_{ck}},{y_{ck}})$; location of covered user points, $({x_i},{y_i}), \forall i \in {\cal{M}}_{{\mathop{\rm cov}} }^k$
\renewcommand{\algorithmicensure}{\textbf{Output:}}
\ENSURE
New horizontal location of aerial BSs, $(x_{ck}^*,y_{ck}^*)$; new radius of coverage areas, $R_k^*$.
\STATE
Find the minimum distance between $({x_{ck}},{y_{ck}})$ and the boundary lines of the corresponding subarea.
\STATE
Obtain $(x_{ck}^*,y_{ck}^*)$ by solving (31).
\STATE
Calculate the minimum distance between $(x_{ck}^*,y_{ck}^*)$ and the boundary lines of the corresponding subarea.
\STATE
Obtain $R_k^*$ from (32).
\end{algorithmic}
\end{algorithm}
It can be seen that, user points which are closer to the centers of aerial BSs have better immunity to estimation errors. The coverage probability decreases when the users which are considered to be covered are actually out of the coverage range of the corresponding aerial BSs.\\
\indent It is clear that increased robustness against imperfect ULI can be achieved by increasing the radii of coverage areas. If we assume the maximum deviation between estimated location and real location for any user point is ${d_{th}}$, where ${d_{th}} \approx 3\sigma$, the performance loss of coverage probability for the $k$-th aerial BS can be completely compensated when 
\begin{equation}
    L_{\min }^k = \left| {{R_k} - {r_{ik}}} \right| \ge {d_{th}}, \forall i \in {\cal{M}}_{{\mathop{\rm cov}} }^k
\end{equation}
 where $L_{\min }^k$ denotes the minimum difference between ${r_{ik}}$ and ${R_k}$ of the $k$-th subarea. There is clearly a trade-off between robustness and required transmit power with regard to the radius. Therefore, the aim of the robust design, which is maximizing the number of covered user points in the existence of inaccurate ULI, is equivalent to maximizing $L_{{\rm{min}}}^k$ with minimum transmit power.  We note that the user points are usually distributed unevenly within the corresponding coverage area for both SD-KM and SD-KMVR techniques. This causes a part of user points having a much larger ground distance to the aerial BS than the rest of user points. Therefore, shifting the circle center to minimize the maximum ground distance between user points covered by the $k$-th aerial BS and the $k$-th circle center can reduce the resulting radius and thus reduce the required transmit power. We denote the distance between $({x_{ck}},{y_{ck}})$ and the boundary lines of the corresponding subarea as ${d_{kl}},l = 1,2,...,{S_k}$. Then the minimum value among all the ${S_k}$ distances is $d_{\min }^k = \min ({d_{kl}})$. 
\begin{figure}
\centering
\subfigure[]{
\label{mosek}
\includegraphics[width=1.55in,height=4.5cm]{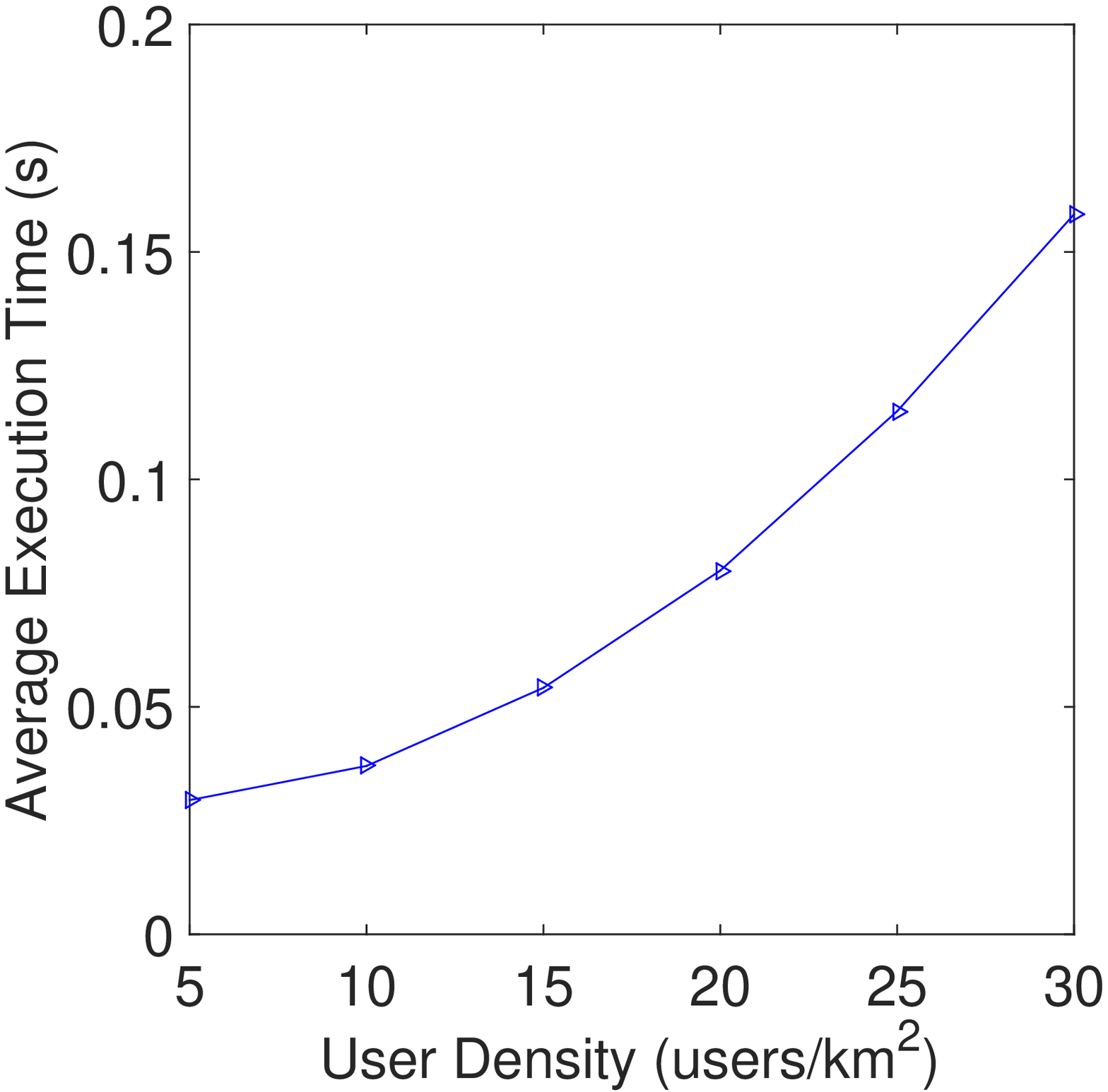}
}
\hspace{-0.1 in}
\subfigure[]{
\label{SD-KMVR}
\includegraphics[width=1.55in,height=4.5cm]{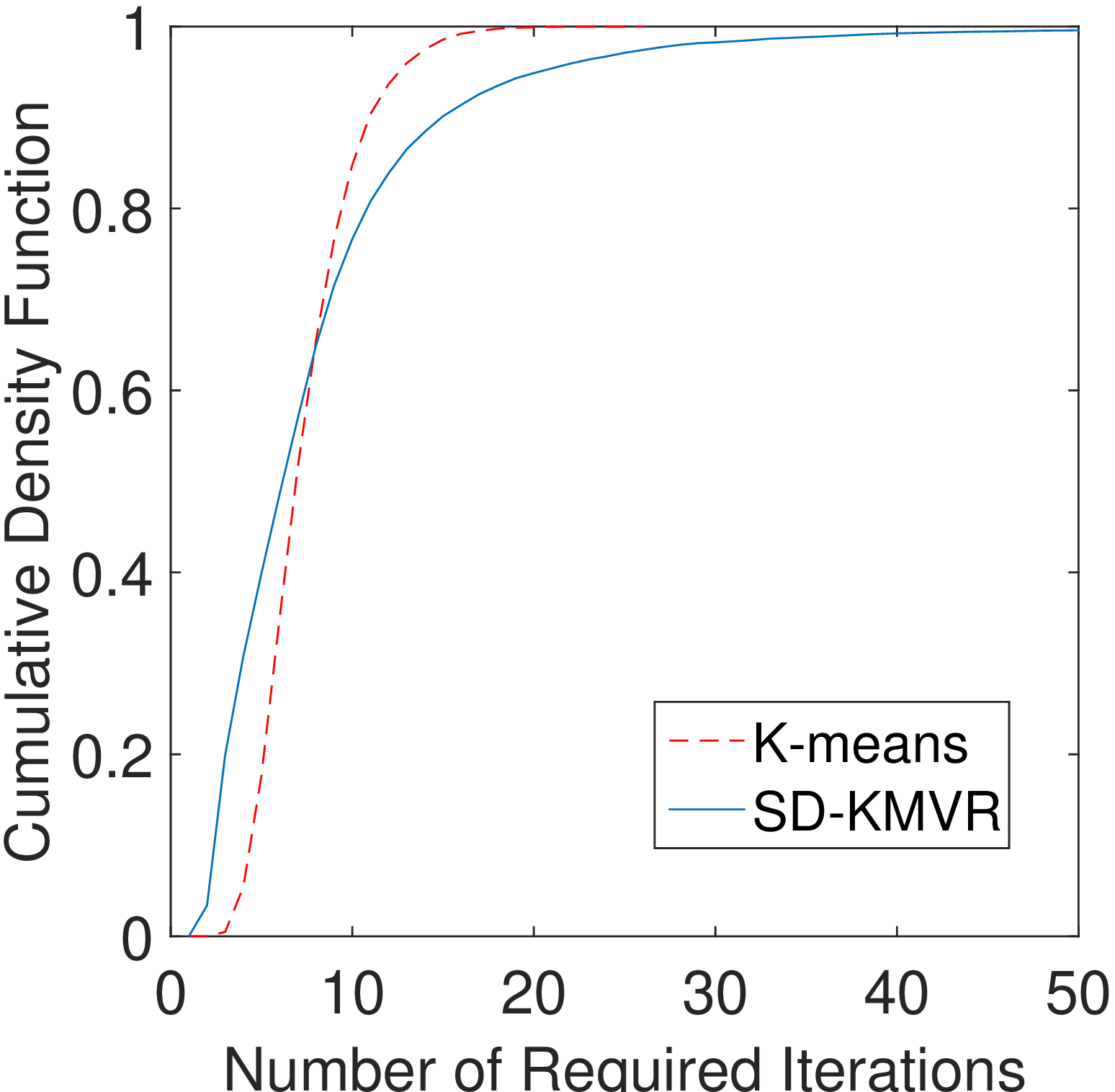}
}
\caption{Computational complexity: (a) average execution time of solving a single MINLP problem by MOSEK solver, $K = 1$; (b) CDF of number of iterations required for $K$-means clustering and SD-KMVR, $K = 9$, ${\lambda _s} = 10$ ${\rm{users/k}}{{\rm{m}}^2}$}
\label{complexity}
\end{figure}
 In order to avoid ICI, the horizontal center of the $k$-th aerial BS can only move within a circular area with radius $d_{\min }^k$. Therefore, the corresponding optimization problem is formulated as
\begin{eqnarray}
    &&\mathop {{\rm{minimize}}}\limits_{x_{ck}^*,y_{ck}^*} {\rm{ }}\mathop {{\rm{max}}}\limits_{i \in M_{{\rm{cov}}}^k} (\sqrt {{{(x_{ck}^* - {x_i})}^2} + {{(y_{ck}^* - {y_i})}^2}} )\\
    &&{\rm{subject}}{\kern 1pt}{\kern 1pt} {\rm{to}}\nonumber\\
    &&\sqrt {{{(x_{ck}^* - {x_{ck}})}^2} + {{(y_{ck}^* - {y_{ck}})}^2}}  \le d_{\min }^k\nonumber\\
    &&k = 1,2,...,K\nonumber
\end{eqnarray}
Here, $(x_{ck}^*,y_{ck}^*)$ is the new horizontal center of the $k$-th aerial BS to optimize, based on the center coordinates $(x_{ck},y_{ck})$ obtained by either SD-KM or SD-KMVR. Note that the objective function of the above optimization problem implicitly includes the constraint that all the originally covered user points are still covered. (30) is equivalent to minimizing an auxiliary variable ${d_k}$ representing the maximum ground distance between $k$-th aerial BS and covered user points according to
\begin{eqnarray}
    &&\mathop {{\rm{minimize}}}\limits_{x_{ck}^*,y_{ck}^*} {\rm{ }}{d_k}\\
    &&{\rm{subject}}{\kern 1pt}{\kern 1pt} {\rm{to}}\nonumber\\
    &&\sqrt {{{(x_{ck}^* - {x_i})}^2} + {{(y_{ck}^* - {y_i})}^2}}  \le {d_k},{\rm{ }}i \in M_{{\rm{cov}}}^k\nonumber\\
    &&\sqrt {{{(x_{ck}^* - {x_{ck}})}^2} + {{(y_{ck}^* - {y_{ck}})}^2}}  \le d_{\min }^k\nonumber\\
    &&k = 1,2,...,K\nonumber
\end{eqnarray}
After obtaining the new center location $(x_{ck}^*,y_{ck}^*)$, we recalculate the minimum ground distance between the $k$-th aerial BS and the corresponding boundary lines and denote it as $d_{\min }^{{k^*}}$. The maximum allowed radius within the $k$-th subarea is then $R_{\max }^{{k^*}} = \min (R,{R_k} + d_{\min }^{{k^*}})$. Therefore, the radius of the $k$-th coverage area $R_k^*$ is 
\begin{equation}  
R_k^*=\left\{            
\begin{array}{lr}  
{{d_k} + {d_{th}},{\rm{ if }}{\kern 1pt}{\kern 1pt}{\kern 1pt}{\kern 1pt}{d_k} + {d_{th}} \le {\rm{R}}_{\max }^{{k^*}}} &\\  
{\rm{R}}_{\max }^{{k^*}},{\rm{if }}{\kern 1pt}{\kern 1pt}{\kern 1pt}{\kern 1pt}{d_k} + {d_{th}} > {\rm{R}}_{\max }^{{k^*}} &    
\end{array}  
\right.  
\end{equation} 
Note that the radius $R_k^*$ is not necessarily larger than ${R_k}$, especially when the technique is applied to SD-KM. Therefore, a reduced transmit power is sometimes obtained. For clarity, the above procedure is summarized in Algorithm 3.

\section{computational complexity analysis}
In this section, we study the computational complexity of the proposed techniques in terms of floating-point operations required. Following \cite{Boyd:2004:CO:993483,8288677}, the computational costs are calculated based on real-valued additions, subtractions, multiplications, divisions and comparisons. 
\subsection{Complexity of SD-GR}
For deploying the $k$-th aerial BS ($k>2$), we first need to find a total of ${{\rm{4}}^{k-1}}$ candidate sets. Each of the ${{\rm{4}}^{k-1}}$ sets is an intersection of $k-1$ sets, and forming each intersection needs $4(k-2)$ comparisons in the worst case for both x and y dimensions. Therefore, the complexity of finding the candidate regions for all the $K$ aerial BSs needs $\sum\limits_{k = 3}^K {2(k - 2){4^k}}$ floating-point operations, which can be simplified as 
\begin{equation}
   C_{GR}^1 = \mathcal{O}\{ \frac{{16K - 14}}{9} \cdot {4^{K + 1}}\}  
\end{equation}
In the elimination process, the complexity arises from the search for feasible regions from all candidate sets. There are a total of $\sum\limits_{k = 2}^K {{4^{K - 1}}}$ candidate sets, and the resulting complexity is obtained as
\begin{equation}
    C_{GR}^2 = \mathcal{O}\{ \frac{1}{3} \cdot {4^{K + 1}}\} 
\end{equation}
The average complexity of solving (18) does not have a closed form solution, and it is in general difficult to determine. Since this complexity is involved in all the deployment schemes, we denote this as ${C_{{\rm{MINLP}}}}$, and represent the complexity of all schemes as a function of this complexity. Therefore, the total computational complexity for SD-GR technique is 
\begin{eqnarray}
    {C_{GR}} &{}={}& {\rm{ }}C_{GR}^{\rm{1}}{\rm{ + }}C_{GR}^2{\rm{ + }}K {C_{{\rm{MINLP}}}}\\
    &{}={}&{\rm{\mathcal{O}\{ }}\frac{{{\rm{6}}K - 11}}{9} \cdot {{\rm{4}}^{K + 1}}{\rm{\}  + }}K {C_{{\rm{MINLP}}}}\nonumber
\end{eqnarray}
To characterize the complexity of solving a single MINLP, instead of an analytical expression, we employ the average execution time against various user density as shown in Fig. 6 (a). 
\subsection{Complexity of SD-KM}
We then consider the complexity of $K$-means clustering. Here, we denote the average number of iterations until convergence as ${n_{{\rm{it}}}}$. The cumulative distribution function (CDF) describing the convergence for $K$-means clustering is shown as the dashed line in Fig. 6 (b). Within each iteration, the proposed scheme involves three steps. The first stage calculates the Euclidean distance between each user point and cluster centers, which involves two subtractions, two multiplications, one addition and one square root. For ${U_{{\rm{tot}}}}$ user points and $K$ clusters, calculating all Euclidean distances requires $\mathcal{O}\{ 6K{U_{{\rm{tot}}}}\}$ floating-point operations. The second stage allocates each user point to the cluster with the closest center, which takes $\mathcal{O}\{ {U_{{\rm{tot}}}}(K - 1)\}$ comparisons. Furthermore, we need to recalculate the cluster centers following (20), which includes ${U_{{\rm{tot}}}}$ multiplications, ${U_{{\rm{tot}}}} - 1$ additions and one division for both x and y dimensions. 
\begin{table}[t!]
\centering
\caption{Computational Complexity of the Proposed Techniques}
\label{complex}
\begin{tabular}{|l|l|}
\hline 
 Method&Computational costs\\
 \hline 
 SD-GR&${\rm{\mathcal{O}\{ }}\frac{{{\rm{6}}K - 11}}{9} \cdot {{\rm{4}}^{K + 1}}{\rm{\}  + }}K {C_{{\rm{MINLP}}}}$  \\
 \hline 
 SD-KM&$\mathcal{O}\{ {N_{{\rm{KM}}}}K{U_{{\rm{tot}}}}n_{{\rm{it}}} + \sum\limits_{k = 1}^K {{{(3 + {S_k})}^{1.5}}} \}+ K {C_{{\rm{MINLP}}}}$ \\
 \hline 
 SD-KMVR&$\mathcal{O}\{ {N_{{\rm{KM}}}}K{U_{{\rm{tot}}}}n_{{\rm{it}}} + \sum\limits_{k = 1}^K {{{(3 + {S_k})}^{1.5}}}$  \\
 &$+ {n_{{\rm{it}}}^{{\rm{vr}}}\sum\limits_{k = 1}^K {{{({U_k} + 1)}^{1.5}}} } \} + (n_{{\rm{it}}}^{{\rm{vr}}} + 1) K {C_{{\rm{MINLP}}}}$\\
 \hline
 Robust&$\mathcal{O}\{ {N_{{\rm{KM}}}}K{U_{{\rm{tot}}}}n_{{\rm{it}}} + \sum\limits_{k = 1}^K {{{(3 + {S_k})}^{1.5}}} \}$ \\
 SD-KM&$+ K {C_{{\rm{MINLP}}}}+\mathcal{O}\{ \sum\limits_{k = 1}^K {{U_k} + K_{}^2} \}$\\
 \hline
 Robust&$\mathcal{O}\{ {N_{{\rm{KM}}}}K{U_{{\rm{tot}}}}n_{{\rm{it}}} + \sum\limits_{k = 1}^K {{{(3 + {S_k})}^{1.5}}}$  \\
 SD-KMVR&$+ {n_{{\rm{it}}}^{{\rm{vr}}}\sum\limits_{k = 1}^K {{{({U_k} + 1)}^{1.5}}} } \} + ({n_{{\rm{it}}}^{{\rm{vr}}}} + 1) K {C_{{\rm{MINLP}}}}$ \\
 &$+\mathcal{O}\{ \sum\limits_{k = 1}^K {{U_k} + K_{}^2} \}$\\
 \hline
\end{tabular}
\end{table}
Then the costs for the third stage is $O\{ 4K{U_{{\rm{tot}}}}\}$. Therefore, the resulting computational complexity of $k$-means clustering is 
\begin{equation}
    \mathcal{O}\{ n_{{\rm{it}}}(6K{U_{{\rm{tot}}}} + {U_{{\rm{tot}}}}(K - 1) + 4K{U_{{\rm{tot}}}})\}  \approx \mathcal{O}\{ K{U_{{\rm{tot}}}}n_{{\rm{it}}}\} 
\end{equation}
To obtain a reasonable value of $K$, the proposed algorithm repeats the above steps for ${N_{{\rm{KM}}}}$ times until $\min (\left\| {{{\bf{m}}_j} - {{\bf{m}}_k}} \right\|) \ge {d_{th}}$ as shown in Section \uppercase\expandafter{\romannumeral4}, and we have
\begin{equation}
    C_{{\rm{KM}}}^1 = \mathcal{O}\{ {N_{{\rm{KM}}}}K{U_{{\rm{tot}}}}n_{{\rm{it}}}\}   
\end{equation}
Moreover, we need to calculate the maximum allowed radius within each cluster according to (23). Following \cite{1709.08278}, (23) is a convex problem solved by interior-point methods with the following complexity:
\begin{equation}
    {C_{{\rm{IP}}}} = \mathcal{O}\{ {(E + F)^{1.5}}{E^2}\} 
\end{equation}
where $E$ is the number of variables, and $F$ is the number of constraints in an optimization problem. As we have a total of ${S_k}$ constraints and 3 variables for solving the $k$-th optimization problem, the computational costs of finding all maximum allowed radius is 
\begin{equation}
    C_{{\rm{KM}}}^2  \approx  \mathcal{O}\{ \sum\limits_{k = 1}^K {{{(3 + {S_k})}^{1.5}}} \}
\end{equation}
The computational complexity of solving (24) is again approximated by ${C_{{\rm{MINLP}}}}$. Accordingly, the total computational cost of SD-KM is
\begin{eqnarray}
    {C_{{\rm{KM}}}} &{}={}& C_{{\rm{KM}}}^1 + C_{{\rm{KM}}}^2 + K{C_{{\rm{MINLP}}}}\\
    &{}={}& \mathcal{O}\{ {N_{{\rm{KM}}}}K{U_{{\rm{tot}}}}n_{{\rm{it}}} + \sum\limits_{k = 1}^K {{{(3 + {S_k})}^{1.5}}} \}   \nonumber\\
    && + K {C_{{\rm{MINLP}}}}\nonumber
\end{eqnarray}
\begin{figure}[t!]
\centering
\includegraphics[width=1\linewidth]{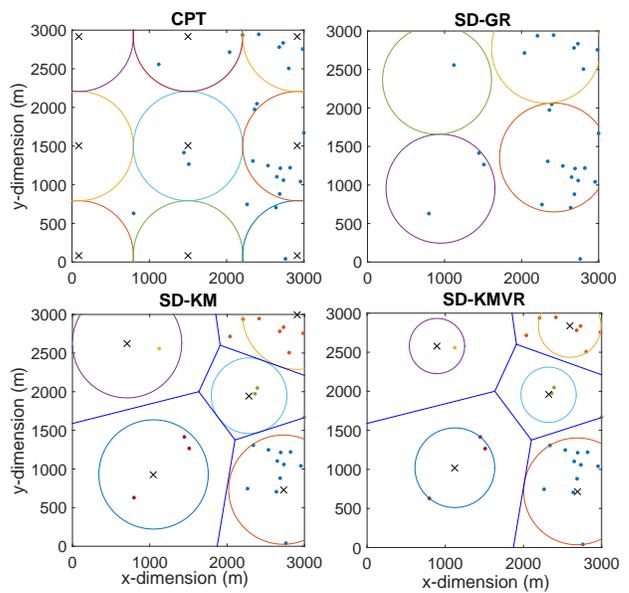}
\caption{Aerial BS placement with proposed techniques}
\label{9}
\end{figure}
\subsection{Complexity of SD-KMVR}
The SD-KMVR technique is based on SD-KM technique and thus involves all operation costs of SD-KM. In addition, SD-KMVR is an iterative algorithm, and denote the average number of iterations required as $n_{{\rm{it}}}^{{\rm{vr}}}$. CDF of the number of required iterations for SD-KMVR technique is also shown in Fig. 6 (b). Within each iteration, the costs of solving (24) is ${C_{{\rm{MINLP}}}}$ and the costs of solving (25) is $\mathcal{O}\{ \sum\limits_{k = 1}^K {{{({U_k} + 1)}^{1.5}}} \}$, since the number of constraints of (25) is ${U_k}$. Therefore, the overall costs of SD-KMVR technique is 
\begin{eqnarray}
    {C_{{\rm{KMVR}}}} &{}={}& {C_{{\rm{KM}}}} + \mathcal{O}\{ n_{{\rm{it}}}^{{\rm{vr}}}\sum\limits_{k = 1}^K {{{({U_k} + 1)}^{1.5}}} \} \\
    && + n_{{\rm{it}}}^{{\rm{vr}}} K {C_{{\rm{MINLP}}}}\nonumber\\
    &{}={}& \mathcal{O}\{ {N_{{\rm{KM}}}}K{U_{{\rm{tot}}}}n_{{\rm{it}}} + \sum\limits_{k = 1}^K {{{(1 + {S_k})}^{1.5}}}+ \nonumber\\
    && {n_{{\rm{it}}}^{{\rm{vr}}}\sum\limits_{k = 1}^K {{{({U_k} + 1)}^{1.5}}} } \} + (n_{{\rm{it}}}^{{\rm{vr}}}+1)K{C_{{\rm{MINLP}}}}\nonumber
\end{eqnarray}
\subsection{Complexity of Robust Technique}
The additional computational costs of applying the robust technique come from (30), which is again solved by interior-point method. Based on (30), $E = 2$ and $F = {U_k} + K - 1$, which leads to
\begin{equation}
    {C_{{\rm{robust}}}}  \approx  \mathcal{O}\{ \sum\limits_{k = 1}^K {{U_k} + K_{}^2} \} 
\end{equation}
\indent For clarity, the computational complexity of the proposed techniques is summarized in Table \ref{complex}. Note that the benchmark CPT is originally designed for maximizing the coverage area instead of maximizing the number of users covered, so the location of aerial BSs is fixed for a specific target area when $R$ is given, and the technique has negligible complexity. To complement the above complexity analysis, in Fig. 11 in the following we show a complexity comparison of the proposed SD-KM, SD-KMVR and robust techniques in terms of average execution time.
\begin{figure}[t!]
\centering
\includegraphics[width=0.95\linewidth,height=6.3cm]{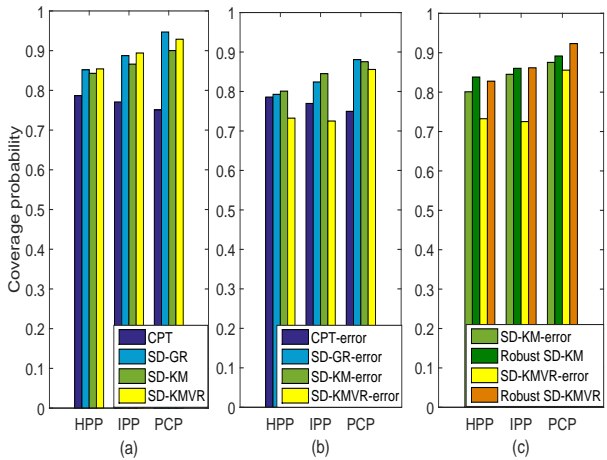}
\caption{User-coverage probability for different types of user distribution: (a) with perfect ULI, (b) with imperfect ULI, (c) with robust technique, $K$=4}
\label{coverage probability}
\end{figure}

\section{simulation results and analysis}
In this section, we assume multiple aerial BSs are vertically deployed at a position which maximizes the coverage area in an urban environment. Therefore, with the parameters shown in Table \ref{my-label}, the radius $R$ corresponds to a path loss threshold ${\rm{P}}{{\rm{L}}_{th}} = 100$ dB is calculated as $R{\rm{ = 707}}$ m. The minimum radius of each coverage area and the minimum allowed distance between clusters are assumed to be ${R_{\min }} = \frac{R}{2}$ and ${d_{th}} = \frac{R}{2}$ respectively. The value of ${K_{\max }}$ is set equal to the number of circles resulting from CPT. We assume ${P_{\min }} =  - 70$ dBm and all the aerial BSs initially have the same transmit power ${P_t} = 30$ dBm. Three different spatial point processes are utilized for modeling the user distribution, which are Homogeneous Poisson process (HPP) with ${\lambda _s} {\rm{ = 5}}$ ${\rm{users/k}}{{\rm{m}}^2}$, Inhomogeneous Poisson process (IPP), with $\lambda (x,y) = 5({x^2} + {y^2})$ ${\rm{users/k}}{{\rm{m}}^2}$ and Poisson cluster process (PCP) respectively. Specifically, the 'parent' points of cluster process are generated following HPP with ${\lambda _p} = 1$ ${\rm{users/k}}{{\rm{m}}^2}$ and the 'children' points are generated with 
\begin{equation}
    {\lambda _c}(x,y) = \frac{\alpha }{{2\pi {\sigma ^2}}}{e^{ - \frac{1}{{2{\sigma ^2}}}({x^2} + {y^2})}}
\end{equation}
where $\alpha  = 0.9$ and $\sigma  = 0.02$. To evaluate the benefit of the proposed algorithms, numerical results based on Monte Carlo simulations of the proposed SD-GR, SD-KM, SD-KMVR and the robust techniques are compared with the performance of CPT which serves as the benchmark. 

\begin{figure}
\centering
\includegraphics[width=0.35\linewidth,height=6.5cm]{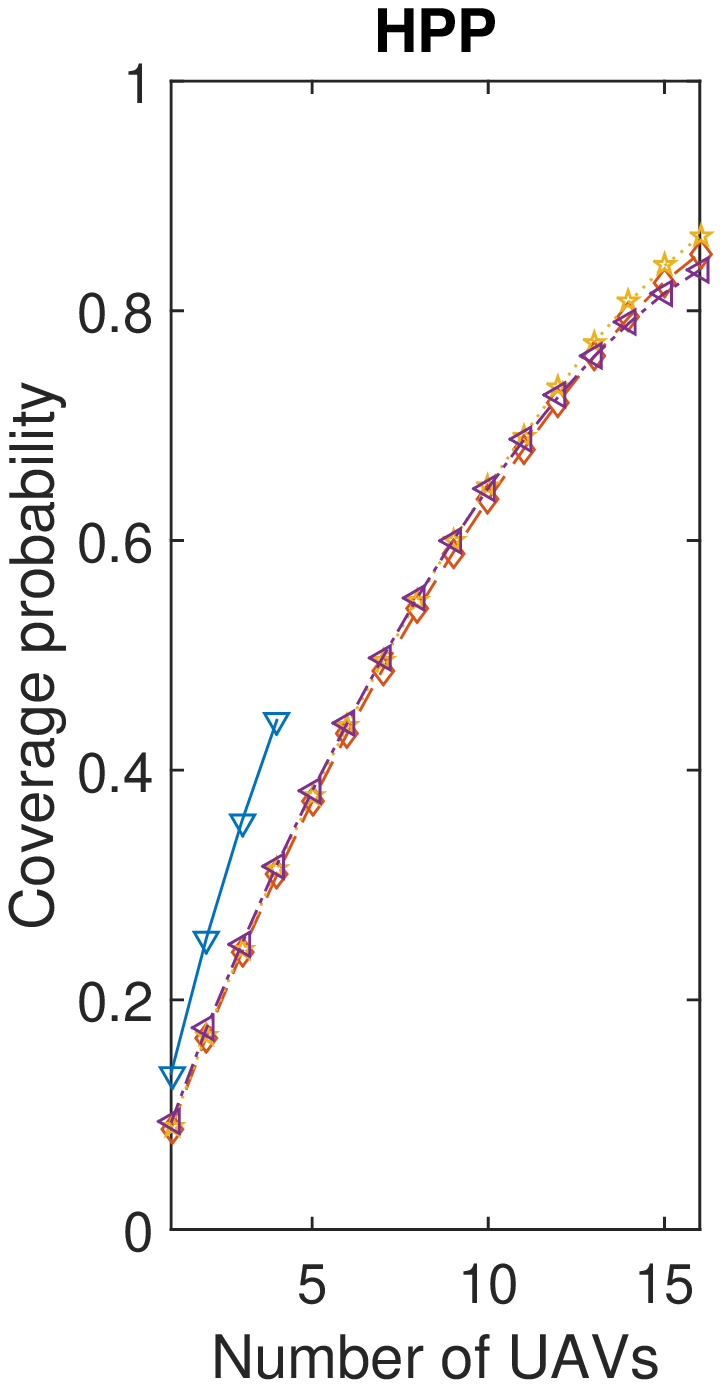}   
\hspace{0.0in}
\includegraphics[width=0.30\linewidth,height=6.5cm]{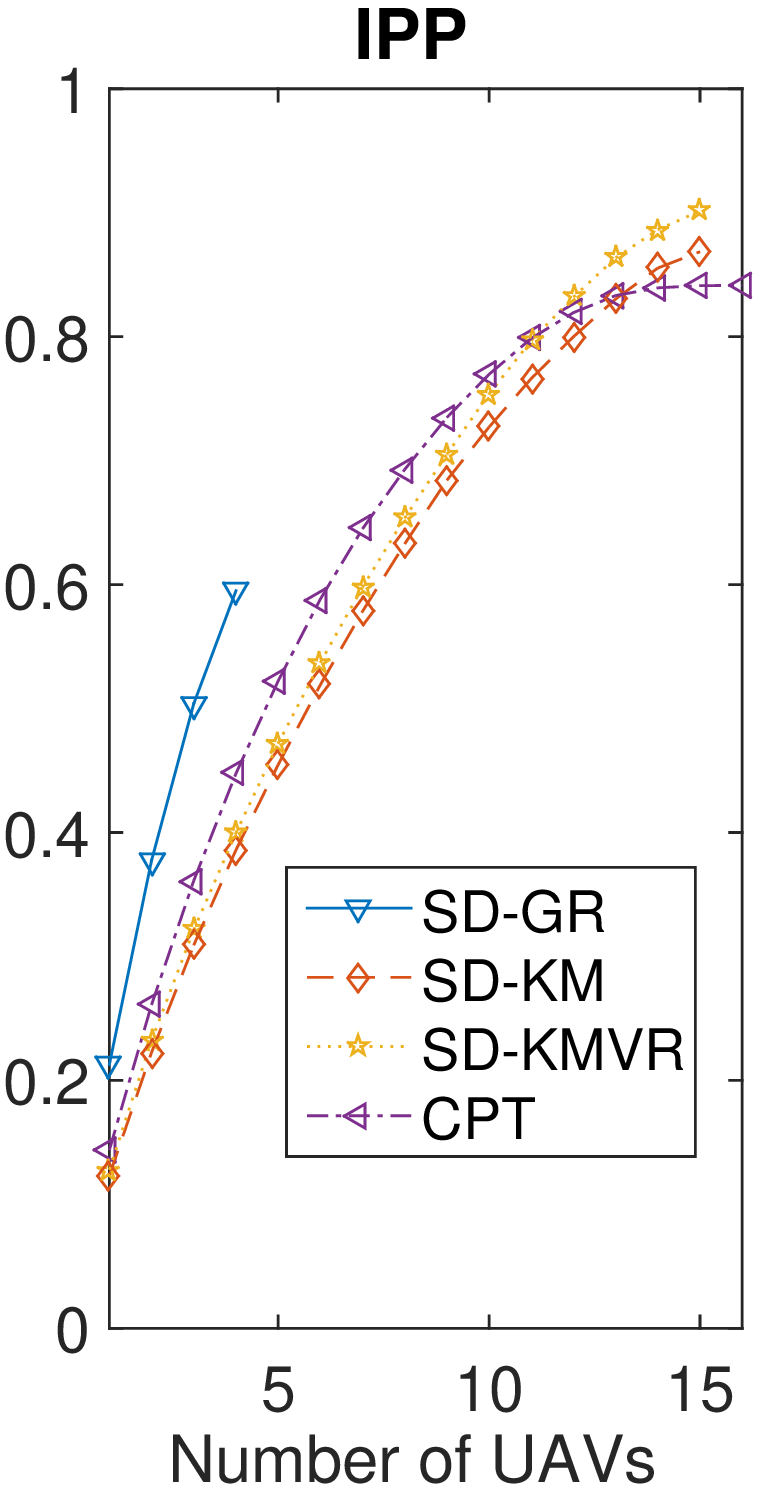}
\hspace{0.0in}
\includegraphics[width=0.30\linewidth,height=6.5cm]{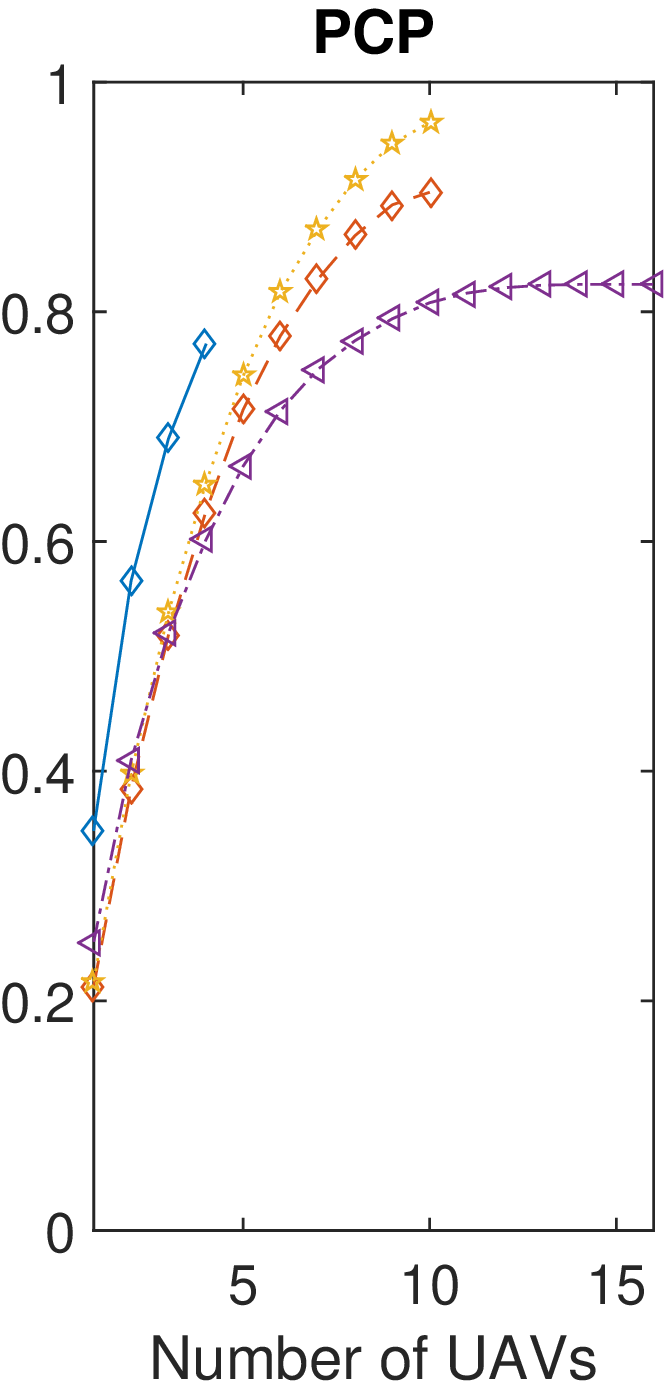}
\caption{User-coverage probability versus number of UAVs deployed, $K$=16, 15 and 10 for HPP, IPP and PCP correspondingly}
\label{limited UAV}
\end{figure}
\begin{table}[t!]
\centering
\caption{Simulation parameters}
\label{my-label}
\begin{tabular}{|l|l|}
\hline 
 parameter&value\\
 \hline 
 $a$&9.61  \\
 \hline 
 $b$&0.16   \\
 \hline 
 ${\eta _{{\rm{LoS}}}}$&1  \\
 \hline
 ${\eta _{{\rm{NLoS}}}}$&20 \\
 \hline
 ${f_c}$&2.5Ghz \\
 \hline
 $c$&$3 \cdot {10^8}$ m/s \\
 \hline
 ${\theta _{opt}}$ & ${42.44^ \circ }$ \\
 \hline
\end{tabular}
\end{table}
Note that due to the exponentially increasing computational complexity of SD-GR as shown in Section \uppercase\expandafter{\romannumeral7}, the maximum $K$ value we use for the SD-GR algorithm is four. The horizontal center of all deployed UAVs must fall inside the target area, and we assume the coverage areas outside the target area will not cause further interference to users outside the interested region.\\ 
\indent To illustrate the function of the proposed solutions, example drone placement distributions are shown based on simulation for the benchmark CPT, SD-GR, SD-KM, and SD-KMVR in Fig. 7, assuming a PCP distribution of users and ${L_s} = 3$ km. The benchmark CPT simply places circles with same size in a way that maximum coverage is achieved and none of these circles overlap. For CPT, the number of circles to be placed in a square target area ${N_{cp}}$ depends on the size of target area, which is ${N_{cp}} = {\left\lceil {\frac{{{L_s}}}{{{\rm{2}}R}}} \right\rceil ^{\rm{2}}}$，where $\left\lceil . \right\rceil $ is a ceiling function. 
\subsection{Coverage Probability}
Intuitively, it can be seen that, coverage probability highly depends on the user distribution when CPT is applied. In addition, SD-GR method always aims to cover the most number of users in the remaining region but the placement of the UAVs is restricted by the previously deployed BSs, which limits the achievable performance. The SD-KM method aims to find the clustering properties among user points and is thus more robust to the change of user distributions. However, the shape of subareas limits the movement of UAVs within each Voronoi cell, which may lead to users gathering in a relative narrow region of the subarea uncovered. This drawback is solved by applying SD-KMVR, which also reduces the transmit power to a large extent.
\begin{figure}[t!]
\centering
\includegraphics[width=1\linewidth,height=6cm]{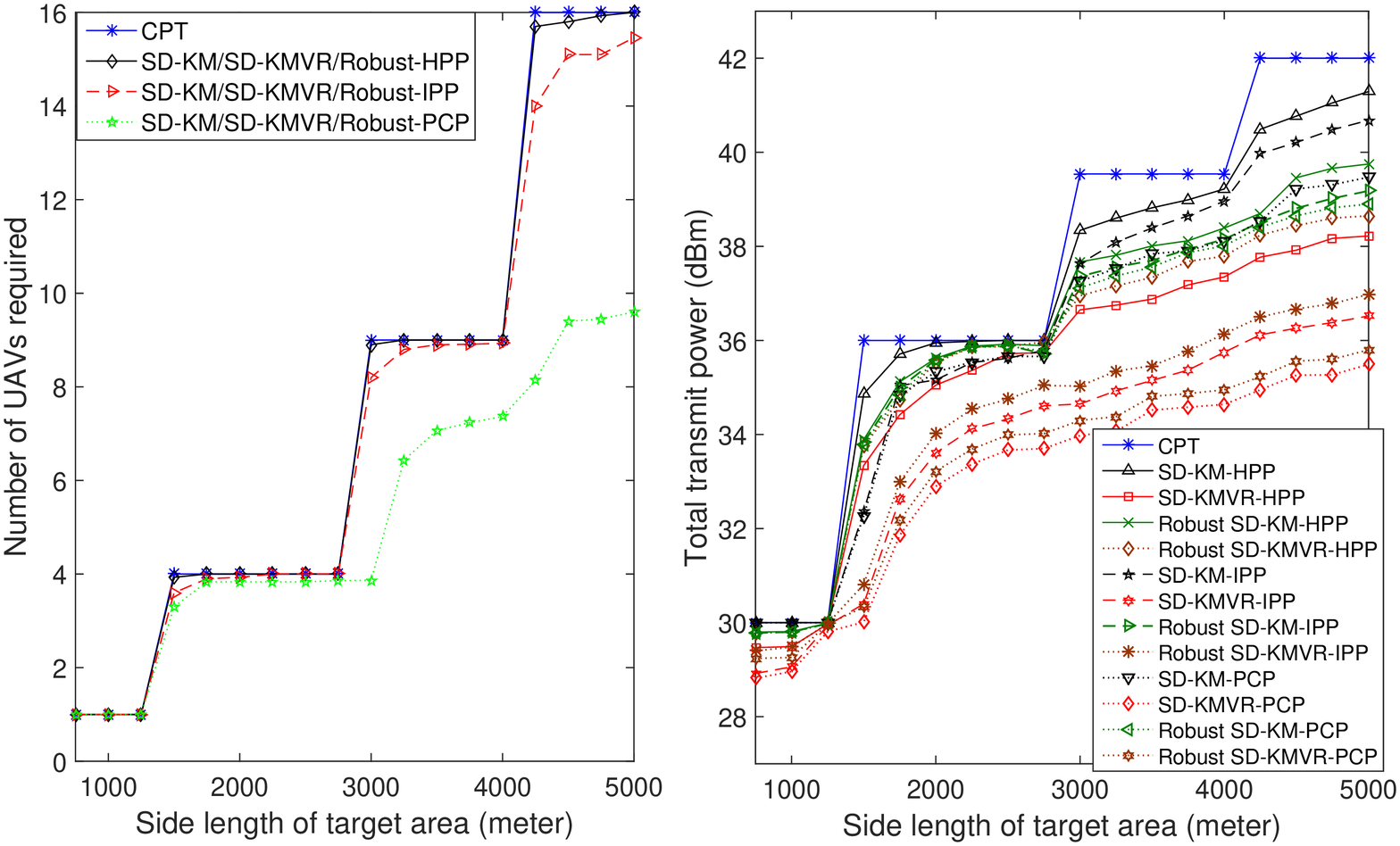}
\caption{Required number of aerial BSs and total transmit power versus size of target area}
\label{13}
\end{figure}
\indent The above effects, are captured in Fig. 8 (a), which illustrates the achieved user-coverage probability, for different types of user distributions. For a fair comparison of the achievable coverage probability, a target area with ${L_s} = 4R$ is assumed, within which all four methods can horizontally deploy a maximum of four circles. It can be observed that the coverage probability of all techniques are affected by user distributions. Note that the performance of CPT decreases while the achieved coverage probability of all the other techniques increase when the user points tend to have an uneven distribution, especially when clusters are formed. Specifically, the proposed SD-KMVR technique achieves an up to 30\% higher coverage probability than the benchmark. The result is expected, because CPT place circles in fixed locations for a given target area no matter how the users are distributed, which highly deteriorate the coverage performance when clusters are formed outside the coverage areas. On the contrary, our proposed algorithms are not restricted to fixed locations, but instead can be flexibly placed according to the change of user distribution. When heterogeneity of user distribution is introduced, especially when clusters are formed, user points are located closer to each other and there is correspondingly a better chance to cover more users within each applied circles. More than 90 percentage of users are covered by applying the proposed techniques when users are distributed following PCP.\\
\indent The coverage performance of the proposed techniques with imperfect ULI is shown in Figure. 8 (b). It can be seen that the performance of all techniques except CPT decreases when introducing imperfect ULI. The performance of CPT remains unchanged because the placement rule of CPT is irrelevant to ULI. Note that SD-KM method shows much better immunity to imperfect ULI than SD-KMVR. This is as expected, since SD-KM method utilizes a larger coverage area causing a larger distance between user points and the border of circles than SD-KMVR. The performance loss is greatly compensated when the proposed robust technique is applied as shown in Figure. 8 (c). The increased coverage probability is achieved as a result of increasing coverage area as well as relocating the aerial BSs.\\
\indent In real scenarios, only a limited number of UAVs may be available for deployment. Therefore, it is meaningful to examine the coverage probability of the proposed techniques versus the number of available aerial BSs. We assume ${L_s} = 5$ km, the $K$ value used for HPP, IPP and PCP are 16, 15 and 10 respectively, which are the average $K$ values we need for target areas of this size, and the corresponding results are shown in Fig. 9. As expected, no matter what user distribution is, the proposed SD-GR technique significantly outperforms other techniques since the UAVs are always deployed in a position such that a maximum number of remaining user points are covered. Moreover, it can be seen that SD-KMVR achieves an up to 10\% performance gain compared to SD-KM. The SD-KM, SD-KMVR and the CPT techniques have comparable performance when users are distributed uniformly. This is because when users are uniformly distributed, the $K$-means clustering method will divide the target area in a similar way as we using CPT. When users are distributed following a non-uniform distribution, CPT outperforms the proposed SD-KM and SD-KMVR methods when only a small number of UAVs are available.  In this case, a similar number of UAVs are deployed in a more tight way than the circle packing technique, causing reduced coverage areas of certain aerial BSs and hence a smaller number of covered users within these areas. When CPT is utilized, circles placed at positions where the user points are densely located can thus cover more user points due to the larger coverage area. However, when clusters are present, the proposed SD-KM and SD-KMVR technique regain the superiority as the UAVs are deployed at positions where clustering properties are found.
\begin{figure}[t!]
\centering
\subfigure[]{
\label{with user}
\includegraphics[width=1.6in,height=5cm]{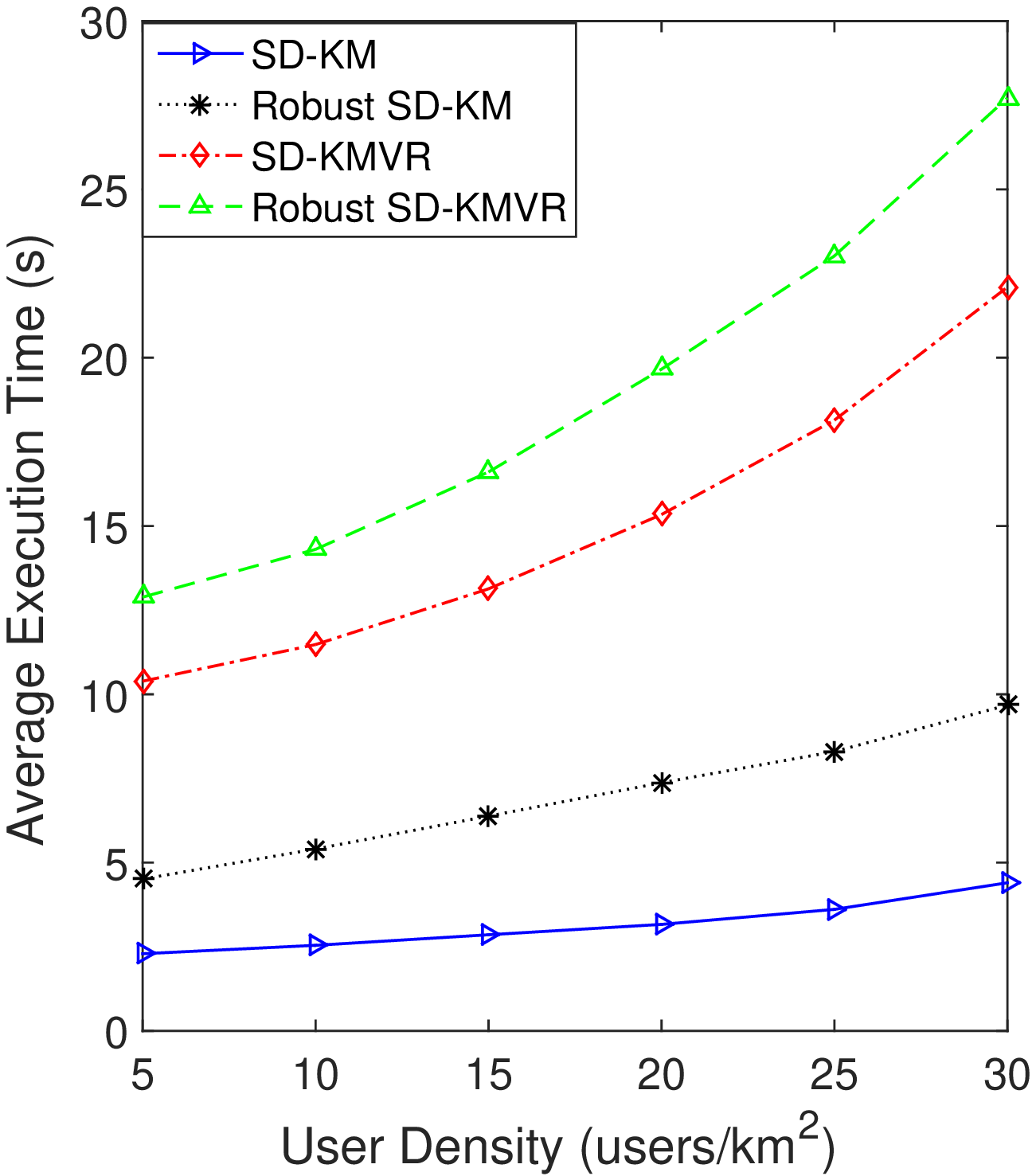}
}
\hspace{-0.2 in}
\subfigure[]{
\label{with UAV}
\includegraphics[width=1.6in,height=5cm]{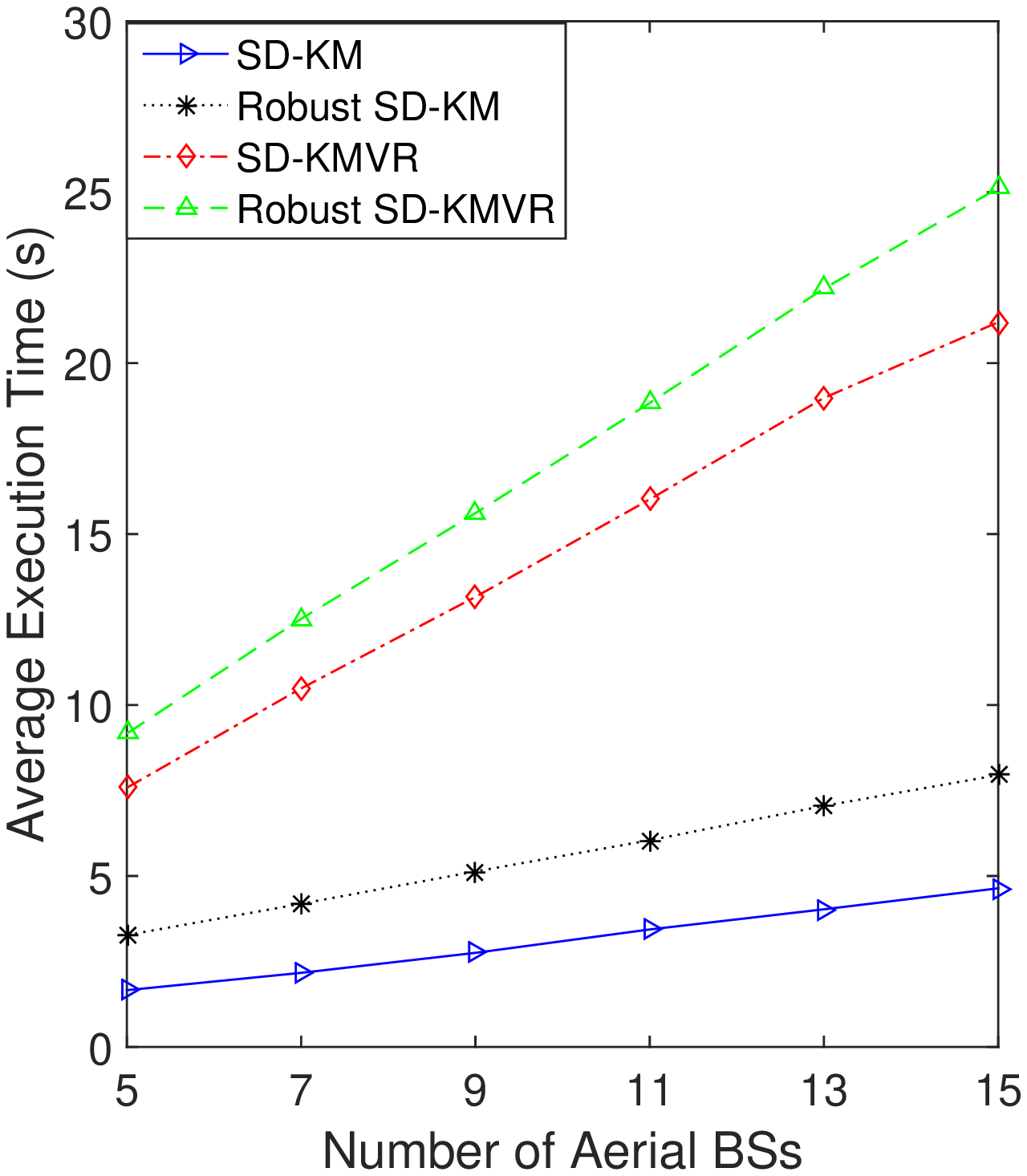}
}
\caption{Average execution time for the proposed techniques: (a) versus various user density, $K$=9 (b) versus various number of aerial BSs, ${\lambda _s} = 5$ ${\rm{users/k}}{{\rm{m}}^2}$  }
\label{execution time}
\end{figure}
\subsection{Energy Efficiency}
In Fig. 10, we compare the number of required aerial BSs and the required total transmit power for SD-KM, SD-KMVR, Robust SD-KM, Robust SD-KMVR and CPT. It can be seen that, SD-KM and SD-KMVR as well as their robust techniques require a smaller number of UAVs compared to CPT when users are not uniformly distributed in the target area. Note that the gap between the proposed algorithms and CPT becomes larger as the size of the target area increases, with SD-KM and SD-KMVR reducing the number of UAVs required down to 60\% of that for CPT. An up to 15\% power is saved by applying SD-KMVR when users are distributed following PCP. In addition, though SD-KM and SD-KMVR require the same number of aerial BSs to be deployed in order to maximize the coverage probability, the SD-KMVR technique saves up to 10\% power as can be seen. Robust SD-KMVR consumes approximately 1\% more power than SD-KMVR as a result of increasing the coverage area, but a clearly reduced transmit power is still obtained compared to SD-KM technique. It is worth highlighting that, Robust SD-KM is even more power efficient than SD-KM, which means the minimum distance between covered user points and the border of the corresponding coverage area after relocating the circle center is larger than ${d_{th}}$ in most cases. Indeed, the reduced number of UAVs saves operation costs and the reduced transmit power can prolong the operation time of aerial BSs. 
\subsection{Computational Complexity}
In Fig. 11, we characterize the complexity of SD-KM, SD-KMVR and their robust techniques in terms of the average execution time. The user points are distributed following HPP and an Intel Core i7-6700 2.6GHz CPU computer is used for performing the simulation. The average execution time versus various user densities is shown in Fig. 11(a) with $K$=9, while the average execution time versus various number of aerial BSs is presented in Fig. 11(b) with ${\lambda _s} = 5$ ${\rm{users/k}}{{\rm{m}}^2}$. It can be observed from both figure that, SD-KMVR technique takes more execution time than SD-KM, and the complexity of SD-KMVR increases more faster. To be specific, the execution time of SD-KMVR increases approximately 40\% and 105\% faster than SD-KM, as the number of user points and the number of aerial BSs increase correspondingly. Similar trends is found for Robust SD-KMVR and Robust SD-KM. Note that the robust techniques also introduce increased computational complexity, which is consistent with the analytic results shown in Section \uppercase\expandafter{\romannumeral7}.

\section{Conclusion}
In this paper, we have studied the efficient deployment of multiple aerial BSs in order to maximize the number of covered users while avoiding ICI. A successive deployment method converting each non-convex constraint into four linear constraints is firstly proposed with geometrical relaxation. In order to reduce the computational complexity, we then propose a simultaneous deployment method based on $K$-means clustering, which converts the target area into $K$ convex subareas and find the optimal location of UAVs within each subarea. Moreover, an iterative algorithm is utilized to reduce the required transmit power while further improve the coverage probability. To increase the robustness against imperfect ULI, a robust technique which relocates the aerial BSs before increasing the radius of coverage areas is proposed. Simulation results show that the proposed algorithms increase the coverage performance by up to 30\%. In addition, SD-GR method is suitable for scenarios where a small number of UAVs are available, and the iterative algorithm achieves an up to 15\% improved power efficiency at a cost of increased computational complexity. Performance loss in the existence of inaccurate ULI is also greatly compensated with the proposed robust technique.

% if have a single appendix:
%\appendix[Proof of the Zonklar Equations]
% or
%\appendix  % for no appendix heading
% do not use \section anymore after \appendix, only \section*
% is possibly needed

% use appendices with more than one appendix
% then use \section to start each appendix
% you must declare a \section before using any
% \subsection or using \label (\appendices by itself
% starts a section numbered zero.)
%

%\appendices
%\section{Proof of the First Zonklar Equation}
%Appendix one text goes here.

% you can choose not to have a title for an appendix
% if you want by leaving the argument blank
%\section{}
%Appendix two text goes here.

% use section* for acknowledgment
%\section*{Acknowledgment}

%The authors would like to thank...

% Can use something like this to put references on a page
% by themselves when using endfloat and the captionsoff option.
\ifCLASSOPTIONcaptionsoff
  \newpage
\fi

% trigger a \newpage just before the given reference
% number - used to balance the columns on the last page
% adjust value as needed - may need to be readjusted if
% the document is modified later
%\IEEEtriggeratref{8}
% The "triggered" command can be changed if desired:
%\IEEEtriggercmd{\enlargethispage{-5in}}

% references section

% can use a bibliography generated by BibTeX as a .bbl file
% BibTeX documentation can be easily obtained at:
% http://mirror.ctan.org/biblio/bibtex/contrib/doc/
% The IEEEtran BibTeX style support page is at:
% http://www.michaelshell.org/tex/ieeetran/bibtex/
%\bibliographystyle{IEEEtran}
% argument is your BibTeX string definitions and bibliography database(s)
%\bibliography{IEEEabrv,../bib/paper}
%
% <OR> manually copy in the resultant .bbl file
% set second argument of \begin to the number of references
% (used to reserve space for the reference number labels box)
%\begin{thebibliography}{1}

%\bibitem{IEEEhowto:kopka}
%H.~Kopka and P.~W. Daly, \emph{A Guide to \LaTeX}, 3rd~ed.\hskip 1em plus
%  0.5em minus 0.4em\relax Harlow, England: Addison-Wesley, 1999.

%\end{thebibliography}
\bibliographystyle{IEEEtran}
\bibliography{list}
%\printbibliography
% biography section
% 
% If you have an EPS/PDF photo (graphicx package needed) extra braces are
% needed around the contents of the optional argument to biography to prevent
% the LaTeX parser from getting confused when it sees the complicated
% \includegraphics command within an optional argument. (You could create
% your own custom macro containing the \includegraphics command to make things
% simpler here.)
%\begin{IEEEbiography}[{\includegraphics[width=1in,height=1.25in,clip,keepaspectratio]{mshell}}]{Michael Shell}
% or if you just want to reserve a space for a photo:

%\begin{IEEEbiography}{Michael Shell}
%Biography text here.
%\end{IEEEbiography}

% if you will not have a photo at all:
%\begin{IEEEbiographynophoto}{John Doe}
%Biography text here.
%\end{IEEEbiographynophoto}

% insert where needed to balance the two columns on the last page with
% biographies
%\newpage

%\begin{IEEEbiographynophoto}{Jane Doe}
%% a \vfill before or after them. The appropriate
% use of \vfill depends on what kind of text is
% on the last page and whether or not the columns
% are being equalized.

%\vfill

% Can be used to pull up biographies so that the bottom of the last one
% is flush with the other column.
%\enlargethispage{-5in}

% that's all folks
\end{document}